\documentclass[%
 reprint,
 amsmath,amssymb,
 aps,
floatfix,
]{revtex4-2}

\usepackage{graphicx}
\usepackage{dcolumn}
\usepackage{bm}
\usepackage[colorlinks,allcolors=black]{hyperref}
\usepackage[capitalize]{cleveref}
\usepackage{aas_macros}

\newif\iftrack
\iftrack
\newcommand{\added}[1]{{\bf #1}}
\newcommand{\deleted}[1]{}
\newcommand{\replaced}[2]{{\bf #2}}
\else
\newcommand{\added}[1]{{#1}}
\newcommand{\deleted}[1]{}
\newcommand{\replaced}[2]{{#2}}
\fi

\newcommand{\avg}[1]{\left\langle#1\right\rangle}

\begin{document}

\preprint{es2022sep15\_818}
\title{Leveraging cross-correlations and linear covariance-based filtering \texorpdfstring{\\}{}for line-intensity map reconstructions at linear scales}

\author{Dongwoo T.~Chung}
 \email{dongwooc@cita.utoronto.ca}
\affiliation{%
 Canadian Institute for Theoretical Astrophysics, University of Toronto, 60 St.~George Street, Toronto, ON M5S 3H8, Canada\\}
\affiliation{Dunlap Institute for Astronomy and Astrophysics, University of Toronto, 50 St.~George Street, Toronto, ON M5S 3H4, Canada
}%

\date{\today}

\begin{abstract}
We explore the possible application of linear covariance-based (LCB) filtering to line-intensity mapping (LIM) signal reconstructions. Originally introduced for reconstruction of the integrated Sachs--Wolfe effect in the cosmic microwave background, the LCB filter is an optimal map estimator that extends the simple Wiener filter by leveraging external correlated data. Given a detectable strong LIM--galaxy or LIM--LIM cross power spectrum, we show recovery of high-redshift, large-scale line-intensity fluctuations---even in the presence of bright interloper emission---in simulations of a futuristic [C\textsc{\,ii}] LIM survey as well as simulated future iterations of the CO Mapping Array Project (COMAP). With sufficient galaxy abundances or low LIM survey noise, normalised cross-correlation between the LCB reconstruction and the true signal reaches 70--90\% on large, linear comoving scales corresponding to $k\sim0.1$ Mpc$^{-1}$. This suggests the possible use of such signal reconstructions in astrophysical or cosmological contexts that require identifying the locations of line emissivity peaks and voids, although clear shortcomings exist on smaller scales. The successful application of the LCB filter in simulated LIM contexts highlights the importance of cross-correlations to studies of the reionising and reionised high-redshift universe with LIM and other large-scale structure surveys.
\end{abstract}

\maketitle


\section{\label{sec:intro}Introduction}

Line-intensity mapping (LIM; for recent reviews see Refs.~\cite{LIM2017,LIM2019,BernalKovetz22}) will provide a view of the cosmic web not through resolving populations of galaxies or other discrete sources, but by tracing the spatial-spectral fluctuations of unresolved, integrated emission in a given atomic or molecular line species. The relevant ideas were first developed for tomography of the high-redshift universe with the 21 cm neutral hydrogen line~\citep{Madau97,Chang08}, but recent activity has spread to theoretical explorations and construction of experiments targeting emission in carbon monoxide (CO) and singly ionised carbon ([C\textsc{\,ii}]) lines~\citep{Lidz11,Gong12,Crites14,Silva15,Yue15,Li16,Keating16,Padmanabhan18,Padmanabhan19,Dumitru19,mmIME-ACA,CCATp,SW21a,SW21b,Yang21b,Cleary22,Padmanabhan22,Bethermin22,CONCERTO}. Observations in those lines could probe the cosmic history of star formation and molecular gas and strongly complement other surveys of large-scale structure (LSS), including 21 cm tomography.

Despite the name, line-intensity mapping is not necessarily about high-fidelity mapping of the redshift-space fluctuations in line intensity, but about measuring the statistics of these fluctuations. Most literature has focussed on recovery of the power spectrum. Some work also explores one-point statistics~\citep{Breysse17,Breysse19,Ihle19} and higher-order statistics, although the latter more so in the context of 21 cm intensity mapping~\citep{Majumdar18,BL18,LaPlante20}. Reconstructing the large-scale line-intensity fluctuations themselves has not been as much of a focus of research.

Yet future LIM experiments will reach sensitivities that make such analyses feasible. Such analyses could potentially also be desirable in many contexts, in terms of being able to designate overdensities or peaks for follow-up, probe underdensities or voids, diagnose the environment of the interstellar medium in a global sense, and so on.

In principle the Wiener filter (abbreviated at points in this work as the WF) provides a simple, minimum-variance linear estimator of the original signal, but only provided that there is sufficient prior knowledge of the signal power spectrum and provided that the signal fluctuations dominate over noise at relevant scales. This will often not be the case even after removal of continuum foregrounds. For instance, [C\textsc{\,ii}] surveys will see prominent interloper line emission from lower redshifts (which would be considered part of the noise in this picture). This emission is expected to be so bright that discriminating it from the [C\textsc{\,ii}] signal is a significant portion of the [C\textsc{\,ii}] LIM theoretical literature~\citep{Breysse15,LidzTaylor16,Cheng16,Sun18,Cheng20}.

Previous literature has also contemplated image processing techniques that use convolutional neural networks in order to resolve line confusion and reconstruction (see, e.g., Refs.~\cite{Moriwaki20,Moriwaki21}). But the expected signal is not well understood (which to be fair is what motivates these surveys in the first place). When experiments forecast the detectability of CO and [C\textsc{\,ii}] LIM signal from the epoch of reionisation (e.g., Refs.~\cite{Breysse22,CCATp}), they show models for the power spectrum that span at least one order of magnitude. Inadequate anticipation of signal and noise easily confounds neural networks without appropriate training (see, e.g., Ref.~\cite{Pfeffer19}). Possibly such uncertainty around the signal has also more generally discouraged interest in exploring LIM signal reconstruction.

One area of significant interest for LIM research, on the other hand, has been the use of cross-correlations involving LIM (or other data sets containing unresolved emission) to improve high-redshift astrophysical and cosmological inferences~\citep{VL10,Pullen13,Serra16,Pullen18,Beane19,Sun19,Yang19,Anderson22,Keenan22}, a line of work that traces all the way back in real-world application to the first detections of 21 cm intensity fluctuations in cross-correlation with spectroscopic galaxy surveys~\citep{Chang10,Masui13,Switzer13}. This interest is a natural inclination. LIM surveys and other LSS surveys all trace the same cosmic web while their systematics and noise will often be disjoint, and cross-correlations will yield scientific output not possible with the auto-correlations in isolation. LIM analyses may also exploit cross power spectra between different parts of its data set to improve estimation of what is actually an auto power spectrum (see, e.g., Ref.~\cite{Ihle19}). So in general, cross-correlations are an invaluable asset for LIM surveys.

One may naturally ask after a reconstruction technique that behaves as linearly and understandably as the Wiener filter but also leverages the power of cross-correlations, and this already exists in the form of the linear covariance-based (LCB) filter. First introduced by~\textcite{Barreiro08}, the LCB filter has mostly been used for reconstruction of maps of the integrated Sachs--Wolfe effect, distinguished from the primordial CMB by cross-correlation with large-scale structure~\citep{Barreiro08,ManzottiDodelson14,PlanckISW,Bonavera16,Weaverdyck18}. This work applies the same formalism to mock LIM data and examines resulting reconstructions in simulated CO and [C\textsc{\,ii}] LIM surveys cross-correlated with galaxy surveys or other LIM data.

To be clear, the central contention of this paper is \emph{not} that the LCB filter is necessarily the best estimator to use. Other reconstruction techniques relying on forward modelling, physical priors (e.g., cosmic tidal shear), Bayesian statistics, information theory, or some combination of these are available~\citep{KitauraEnsslin08,Pen12,Opperman13,Zhu16,Modi19}. But for the purposes of this work, we want to demonstrate the extent to which cross-correlations can indeed improve reconstruction of LIM signals relative to relying on auto-correlation alone, with the LCB filter and Wiener filter representing the optimal solutions in the linear regime in each case.

More specifically, we ask: Can the LCB filter successfully reconstruct the structure of line-intensity fluctuations on large linear scales by exploiting
\begin{itemize}
    \item cross-correlations with surveys of discrete sources?
    \item cross-correlation with a measurement of a different, correlated line-intensity signal?
\end{itemize}

The paper is structured as follows. We provide an overview of the LCB filter itself in~\autoref{sec:lcbtop}, reviewing its derivation and considering cases relevant to LIM reconstruction. After a brief interlude in~\autoref{sec:intermezzo} about metrics for `successful reconstruction', we then consider a proof of concept for applying the LCB filter to LIM with simulations of [C\textsc{\,ii}] signal reconstruction in~\autoref{sec:CIILIM}, and undertake additional case studies in the context of the CO Mapping Array Project (COMAP) in~\autoref{sec:COLIM}. We end by drawing conclusions from those results in~\autoref{sec:conclusions}.

We use base-10 logarithms throughout unless otherwise specified, as well as $\Lambda$CDM cosmologies with parameters $\{\Omega_m,\Omega_\Lambda,\Omega_b,h\equiv H_0/(100$\,km\,s$^{-1}$\,Mpc$^{-1}),\sigma_8,n_s\}$ stated near the start of each section as appropriate. Distances carry an implicit $h^{-1}$ dependence throughout, which also propagates through to masses (all based on virial halo masses $\propto h^{-1}$) and volume densities ($\propto h^3$).

\section{The LCB filter}
\label{sec:lcbtop}
We first recap the derivation of~\textcite{Barreiro08}, but rephrased in terms of the isotropic power spectrum $P(k)$ describing a random Gaussian field, instead of specifically in a spherical multipole expansion. This in and of itself is largely a superficial change as mathematically the two phrasings are equivalent. However, we provide the derivation already generalised to reconstructing the desired signal from cross-correlation with $n$ other tracers.

Note also that~\textcite{ManzottiDodelson14} derive an ISW estimator based on likelihood optimisation, but state that it will also reduce to the LCB estimator. We verify this explicitly in~\cref{sec:lcb_md14}. Thus, the LCB filter describes an optimal estimator for the signal that combines cross-correlation and auto-correlation measurements across all observables.

Following the general derivation provided in~\autoref{sec:lcbgen}, we consider in~\autoref{sec:lcbn1} the $n=1$ case originally described by~\textcite{Barreiro08}, and then consider aspects of practical implementation for matter tracers in~\autoref{sec:lcbbias}.

\subsection{General case}
\label{sec:lcbgen}
We begin by noting that the auto and cross power spectra by definition describe the covariance of the Fourier modes of the random Gaussian fields:
\begin{equation}
    P^{(S)}_{ij}(k) = \avg{s_i(k)s_j^*(k)}.
\end{equation}
Note that we have assumed $\avg{s_i(k)} = 0$ for all $k$ and $i$, implying that any non-zero mean that may have existed has already been subtracted in mapping the field. 

Assume mode coupling is negligible, i.e., the covariance matrix of the signal fields is block-diagonal with all cross-$k$ elements equal to zero and described fully by $P^{(S)}_{ij}(k)$. The covariance will also be symmetric, as $P^{(S)}_{ij}(k)=P^{(S)}_{ji}(k)$. Assume furthermore that we obtain measurements $d_i$ of each field with uncorrelated additive Gaussian noise:
\begin{equation}
    d_i=s_i+n_i.
\end{equation}
If the power spectrum of each noise component is thus some constant $P_i^{(N)}$, then the total covariance matrix is given by
\begin{equation}
    C_{ij}(k) = P^{(S)}_{ij}(k) + \delta_{ij}P_i^{(N)},
\end{equation}
with $\delta_{ij}$ being the Kronecker delta. Going forward, we write the auto power spectra as $P_i^{(S)}(k)\equiv P_{ii}^{(S)}(k)$ for brevity.

Then given measurements $\{d_1,d_2,\dots,d_n\}$ of tracers $\{s_1,s_2,\dots,s_n\}$, we wish to obtain a reconstruction of $s_{n+1}$ from data $d_{n+1}$. For brevity we notate $p\equiv n+1$, and we will further abbreviate the signal and noise power spectra for this last observable as $P^{(S)}\equiv P_{p}^{(S)}$ and $P^{(N)}\equiv P_{p}^{(N)}$.

Consider for a moment the covariance of the vector $\mathbf{f}=(d_1,d_2,\dots,d_n,s_{p})$, i.e., the covariance between the observations of all $n$ tracers that we are not reconstructing and the intrinsic signal of the tracer $s_p$ that we do want to reconstruct. This then is given by
\begin{equation}
    D_{ij}(k) = \begin{cases}P^{(S)}_{ij}(k) + \delta_{ij}P_i^{(N)}&\text{if }i\leq n\text{ and }j\leq n,\\P^{(S)}_{ij}(k)&\text{otherwise.}\end{cases}\label{eq:Dijdef}
\end{equation}
with $\delta_{ij}$ being the Kronecker delta.

We are interested in the Cholesky decomposition of this matrix, i.e., the identification of a lower triangular matrix $L(k)$ such that $D=LL^T$ at each $k$. This is because once we obtain $L$, we can claim that some hidden Gaussian random vector $\mathbf{h}$, whose elements are all uncorrelated with diagonal unit covariance, has generated our correlated observations and signals $f_i$ such that
\begin{equation}
    \mathbf{f}(k) = L(k)\mathbf{h}(k).
\end{equation}

What we then want to do is express the signal $s_{p}$ in terms of the hidden $\mathbf{h}$ and the other data $d_i$, as well as the lower triangular matrix $L$. At each $k$,
\begin{align}
    s_{p} &= \sum_{i=1}^{p}L_{pi}h_i\nonumber\\&= L_{pp}h_{p} + \sum_{i=1}^{n}L_{pi}h_i\nonumber\\&= L_{pp}h_{p} + \sum_{i=1}^{n}\left(L_{pi}\sum_{j=1}^n(L^{-1})_{ij}d_j\right),\label{eq:snplus1}
\end{align}
noting that we write the row index first when indexing the matrices. For the last step we recall that $L^{-1}\mathbf{f} = \mathbf{h}$, and note that $L^{-1}$ is also a lower triangular matrix so that $(L^{-1})_{ip}=0$ for all $i\leq n$.

Note that no knowledge of the hidden fields is necessary to calculate the second term, which is purely defined in terms of the data--signal covariances and the data for tracers 1 through $n$. At this point,~\textcite{Barreiro08} propose that to estimate the scaled hidden field $L_{pp}h_{p}$, one should Wiener-filter the difference between the actual data and the second term of~\cref{eq:snplus1}, i.e.,
\begin{align}
    \bar{d}&\equiv d_{p}-\sum_{i=1}^{n}\left(L_{pi}\sum_{j=1}^n(L^{-1})_{ij}d_j\right)\nonumber\\
    &=L_{pp}h_{p} + n_{p}.
\end{align}

Then the linear-covariance based estimator for $s_{p}$ simply substitutes the above (with a Wiener filter factor) into~\cref{eq:snplus1}. With a slight rearrangement of the double sum, we can write the estimator as
\begin{align}
    \hat{s}_{p} &= \frac{L_{pp}^2}{L_{pp}^2+P^{(N)}}\left(d_{p}-\sum_{j=1}^n\sum_{i=1}^{n}L_{pi}(L^{-1})_{ij}d_j\right)\nonumber\\&\quad+\sum_{j=1}^n\sum_{i=1}^{n}L_{pi}(L^{-1})_{ij}d_j.\label{eq:LCBinitial}
\end{align}
This is analogous to the expression given by Refs.~\cite{PlanckISW,Bonavera16}.

The form of~\cref{eq:LCBinitial} also allows us to easily see the power spectrum of the reconstructed $\hat{s}_p$. By definition, $\bar{d}$ is uncorrelated with the second term of~\cref{eq:snplus1}, and should have a power spectrum given by $L_{pp}^2+P^{(N)}$. The power spectrum of all of $\hat{s}_p$ should then be
\begin{align}
    P^{(\hat{s})} &= \frac{L_{pp}^4}{L_{pp}^2+P^{(N)}}\nonumber\\&\quad+\sum_{j=1}^n\sum_{j'=1}^n\sum_{i=1}^{n}\sum_{i'=1}^{n}L_{pi}L_{pi'}(L^{-1})_{ij}(L^{-1})_{i'j'}D_{jj'}.\label{eq:LCBPk}
\end{align}

However,~\cref{eq:LCBinitial} can also be clearly rewritten as
\begin{align}
    \hat{s}_{p} &= \frac{L_{pp}^2}{L_{pp}^2+P^{(N)}}d_{p}\nonumber\\&\quad+\left(1-\frac{L_{pp}^2}{L_{pp}^2+P^{(N)}}\right)\sum_{j=1}^n\sum_{i=1}^{n}L_{pi}(L^{-1})_{ij}d_j.\label{eq:LCB}
\end{align}
This provides an alternate intuition for what is going on: the estimator starts with filtered data for the observable $p$, and then introduces additional fluctuations based on correlations between the observables. Indeed, with $n=0$ (i.e., without anything to cross-correlate with $d_{p}$) we would simply recover the Wiener filter. However, to delve further into this kind of intuition, it is useful to return to the context of the original derivation given by~\textcite{Barreiro08}, where the signal to be reconstructed is only correlated with one additional tracer.

\subsection{Special case---\texorpdfstring{$n=1$}{n=1}}
\label{sec:lcbn1}
Suppose we only have two observables, $d_1$ and $d_2$, and we wish to obtain the estimator $\hat{s}_2$ for the signal underlying the latter observation. In this case, the upper triangular matrix $L(k)$ from the Cholesky decomposition is straightforwardly found at each $k$:
\begin{align}
    L_{11}&=\left(P_1^{(S)}+P_1^{(N)}\right)^{1/2};\label{eq:chol11n1}\\L_{21}&=P_{12}^{(S)}\left(P_1^{(S)}+P_1^{(N)}\right)^{-1/2};\label{eq:chol12n1}\\L_{22}&=\left(P^{(S)}-\frac{{P_{12}^{(S)}}^2}{P_1^{(S)}+P_1^{(N)}}\right)^{1/2};\label{eq:chol22n1}
\end{align}
and of course $L_{12}=0$. General, recursive expressions to evaluate $L$ from $D$ exist for $n>1$ also, and are given in~\cref{sec:lcbmaths}.

As far as the inverse of this matrix is concerned, only the element $(L^{-1})_{11}$ will be relevant in evaluating~\cref{eq:LCB} given that $n=1$, and it is extremely straightforward to find:
\begin{align}
    (L^{-1})_{11}&=1/L_{11}=\left(P_1^{(S)}+P_1^{(N)}\right)^{-1/2}.
\end{align}
Substituting the above into~\cref{eq:LCB} yields
\begin{align}
    \hat{s}_{2} &= \frac{L_{22}^2}{L_{22}^2+P^{(N)}}d_{2}+\left(1-\frac{L_{22}^2}{L_{22}^2+P^{(N)}}\right)\left(\frac{L_{12}}{L_{11}}d_1\right)\nonumber\\&= \frac{L_{22}^2}{L_{22}^2+P^{(N)}}d_{2}\nonumber\\&\quad+\left(1-\frac{L_{22}^2}{L_{22}^2+P^{(N)}}\right)\frac{P_{12}^{(S)}}{P_1^{(S)}+P_1^{(N)}}d_1,\label{eq:LCBn1}
\end{align}
where
\begin{equation}
\frac{L_{22}^2}{L_{22}^2+P^{(N)}}=\frac{\left(P_1^{(S)}+P_1^{(N)}\right)P^{(S)}-{P_{12}^{(S)}}^2}{\left(P_1^{(S)}+P_1^{(N)}\right)\left(P^{(S)}+P^{(N)}\right)-{P_{12}^{(S)}}^2}.\label{eq:LCBn1ratio}
\end{equation}
It is interesting to consider some limiting cases of this. If the noise in $d_1$ is overwhelmingly large, then the weights for $d_1$ go to zero and the best estimator for $s_2$ is simply the Wiener-filtered data $d_2$. On the other hand, if the noise in $d_2$ is overwhelmingly large (i.e., $P^{(N)}\gg L_{22}^2$), then the weights for $d_2$ go to zero instead so that the best estimator for $s_2$ is actually simply a filtered version of $d_1$, except the weights have the cross power spectrum $P_{12}^{(S)}$ in the numerator where the Wiener filter weights would have had $P_1^{(S)}$. Where $\hat{s}_2$ lands between these extremes simply depends on the relative signal and noise power spectrum amplitudes.

The power spectrum of the estimator is given by setting $n=1$ in~\cref{eq:LCBPk}:
\begin{align}
    P^{(\hat{s})} &= \frac{\left[\left(P_1^{(S)}+P_1^{(N)}\right)P^{(S)}-{P_{12}^{(S)}}^2\right]^2/\left(P_1^{(S)}+P_1^{(N)}\right)}{\left(P_1^{(S)}+P_1^{(N)}\right)\left(P^{(S)}+P^{(N)}\right)-{P_{12}^{(S)}}^2}\nonumber\\&\quad+\frac{{P_{12}^{(S)}}^2}{P_1^{(S)}+P_1^{(N)}}.\label{eq:LCBPkn1}
\end{align}
The values of this expression in the limit of large $P_1^{(N)}$ or large $P^{(N)}$ are each consistent with expectations for power spectra of Wiener-filtered data.

As with the Wiener filter, the LCB filter suppresses fluctuations relative to the original signal:
\begin{align}
    P^{(S)}-P^{(\hat{s})} &=
    \frac{\left[(P_1^{(S)}+P_1^{(N)})P^{(S)}-{P_{12}^{(S)}}^2\right]P^{(N)}}{\left(P_1^{(S)}+P_1^{(N)}\right)\left(P^{(S)}+P^{(N)}\right)-{P_{12}^{(S)}}^2}.\label{eq:LCBPkn1bias_interim}
\end{align}
The cross power spectrum magnitude cannot exceed the geometric mean of the total auto power spectra, i.e., we can define
\begin{equation}
    \alpha\equiv\frac{\left|P_{12}^{(S)}\right|^2}{\left(P_1^{(S)}+P_1^{(N)}\right)\left(P^{(S)}+P^{(N)}\right)},
\end{equation}
which must satisfy $0\leq\alpha\leq1$. Then
\begin{align}
    P^{(S)}-P^{(\hat{s})} &=
    \frac{\left(P^{(S)}-\frac{\alpha}{1-\alpha} P^{(N)}\right)P^{(N)}}{P^{(S)}+P^{(N)}}.\label{eq:LCBPkn1bias}
\end{align}
In the case of $\alpha=0$, where there is no correlation between observables 1 and 2, we simply recover the power spectrum bias of the Wiener filter for observable 2 alone. However, it is clear that non-zero correlation (i.e., non-zero values of $\alpha$) would reduce the bias down from this limit, potentially significantly depending on the relative amplitudes of the signal and noise power spectra. This would suggest that cross-correlations allow recovery of additional fluctuations in the underlying signal.

\subsection{Special case---\texorpdfstring{$s_i=B_i s$}{si=Bis} for all \texorpdfstring{$i$}{i}}
\label{sec:lcbbias}

Suppose all our observations are of the same underlying signal, but subject to a different linear bias. One might consider, for example, that fluctuations in line brightness temperature $\delta T_\text{line}$ trace the matter density contrast $\delta_m$ scaled by the luminosity-weighted average temperature--bias product $\avg{Tb}_\text{line}$. In a halo model that relates halo mass $M_h$ and redshift $z$ to a line luminosity $L(M_h,z)$, we may calculate this quantity as
\begin{align}
    \avg{Tb}_\text{line} &= \frac{c^3(1+z)}{8\pi k_B\nu_{\text{rest}}^3 H(z)}\nonumber\\&\qquad\times\int dM_h\,\frac{dn}{dM_h}\,L(M_h,z)\,b(M_h),\label{eq:Tbdef}
\end{align}
given the linear bias $b(M_h)$ for halos of virial mass $M_h$, the halo mass function $dn/dM_h$, the rest-frame line emission frequency $\nu_\text{rest}$, the Hubble parameter $H(z)$ for the assumed cosmology, the speed of light $c$, and the Boltzmann constant $k_B$. (When working in units of flux density rather than brightness temperature, replacing the multiplier in front of the integral with simply $c/[4\pi\nu_\text{rest}H(z)]$ in~\cref{eq:Tbdef} gives the mean intensity--bias product. See, e.g., Ref.~\citep{selfcite_lims3}.)

Then on large linear scales, the matter power spectrum $P_m$ and the line-intensity power spectrum $P_\text{line}$ are related by
\begin{equation}
    P_\text{line}(k)\simeq\avg{Tb}_\text{line}P_m(k),
\end{equation}
where we ignore the Poissonian shot noise that begins dominating at nonlinear scales. This is hardly a novel insight, and the vast majority of LIM forecasts rely on halo models with linear bias. Galaxy number density contrast as measured by a galaxy survey will of course have a similar effective bias $b_\text{gal}$ such that the galaxy density contrast has a clustering power spectrum $P_\text{gal}(k)\simeq b_\text{gal}P_m(k)$. The key point is that this bias parameter, be it $\avg{Tb}_\text{line}$ or $b_\text{gal}$, differs across the range of atomic lines, molecular lines, and galaxy selections. If we had only observations of the exact same tracer, the LCB filter would simply reduce to an inverse variance-weighted Wiener filter, which we show explicitly in~\cref{sec:lcbcommon}.

Return to a more general picture for a moment. Instead of dealing specifically with $\delta_m$ and some combination of $\avg{Tb}_\text{line}$ or $b_\text{gal}$ parameters, suppose we have an underlying field $s$ and observable fields $s_i$ for $i\in\{1,\dots,n,p\equiv n+1\}$ all tracing $s$ with some linear bias such that $s_i=B_is$ for all $i$. Then the signal auto and cross power spectra are all proportional to some common power spectrum. Call this $P(k)$, so that $P^{(S)}_{ij}(k)=B_iB_jP(k)$.

As before we would like to go about reconstructing $s_p$. Defining the appropriate matrix $D$, we have
\begin{equation}
    D_{ij}(k) = \begin{cases}B_iB_jP(k) + \delta_{ij}P_i^{(N)}&\text{if }i\leq n\text{ and }j\leq n,\\B_iB_jP(k)&\text{otherwise.}\end{cases}\label{eq:Dijdefs}
\end{equation}
With prior knowledge of $B_p$ and $P(k)$, application of the LCB filter is then straightforward. However, what if we were to plead ignorance of either $B_p$ or $P(k)$? Even in the absence of correlated noise, we would not be able to claim knowledge of $B_p^2P(k)$ unbiased by $P_p^{(N)}$ from one data set $d_p$ by itself.

In the case where $s_i=s_j$ for all $i$ and $j$ (e.g., all $B_i=1$), every cross spectrum is effectively an unbiased estimator of the signal power spectrum. As we mentioned above, \cref{sec:lcbcommon} shows that this case simply reduces to the Wiener-filtered reconstruction of the inverse noise variance-weighted average of $d_i$.

If the signals are biased by factors $B_i$ that are not necessarily equal, estimation of $P^{(S)}$ is less straightforward. In fact, if $n=1$ (i.e., we have two independent observations), we would only have one cross spectrum, and this would not be sufficient for unbiased estimation of $B_p^2P(k)$.

However, $n\geq2$ is sufficient to estimate $B_p^2P(k)$ based on the cross spectra. It is straightforward to see that in the absence of correlated noise, we expect
\begin{equation}
    \hat{P}_p^{(S)}(k)=\frac{2}{n(n-1)}\sum_{i=1}^{n-1}\sum_{j=i+1}^n\frac{P_{ip}^{(S)}(k)P_{jp}^{(S)}(k)}{P_{ij}^{(S)}(k)}.
\end{equation}
Having an estimator in this vein then allows us to define $D$ purely in terms of total auto and cross power spectra. Denoting the total auto power spectrum of $d_i$ as $P^{(T)}_i\equiv P^{(S)}_i+P^{(N)}_i$, we have
\begin{equation}
    D_{ij}(k) = \begin{cases}P^{(T)}_i(k)&\text{if }i=j\leq n,\\P^{(S)}_{ij}(k)&\text{if }i\neq j,\\\hat{P}_p^{(S)}(k)&\text{if }i=j=p.\end{cases}\label{eq:Dijpractical}
\end{equation}
Note that none of this actually requires $B_i\neq B_j$ for all $i$ and $j$. So for instance, we could leverage two independent observations of one line to reconstruct a third independent observation of a different line tracing the same matter density fluctuations.

\section{Intermezzo---quantifying the success of optimal estimators}
\label{sec:intermezzo}
Before moving to consider potential applications of the LCB filter, we need to address how this work will quantify \emph{`successful reconstruction'} of the structure of line-intensity fluctuations.

It is clear that the LCB filter results in optimal estimation of the target signal, but with a power spectrum that is suppressed relative to that of the original signal (although less so than estimation with the Wiener filter). Furthermore, the assumptions behind the design of the LCB filter ultimately almost all run up against the reality of the LIM signal. The signal for many atomic and molecular lines will be intrinsically highly non-Gaussian if originating from later cosmic epochs and principally tracing galaxies (whose statistics have long been understood to behave less like a Gaussian random field and more like Poissonian draws from an underlying lognormal field---see, e.g.,~\textcite{ColesJones91}).

Furthermore, even if the line-intensity field in actual comoving space were a Gaussian random field, what a LIM survey will actually be able to observe is the \emph{pseudo}-intensity field convolved with instrumental response and survey transfer functions. Thus, the output of the LCB estimator will be a \emph{pseudo}-pseudo-intensity field, a suppressed version of the fluctuations already suppressed by the observation.

The central aim of this paper that was stated in the~\hyperref[sec:intro]{Introduction} is not to show that the LCB filter reconstructs the amplitudes of line-intensity fluctuations. Rather, we will show that it allows reconstruction of the \emph{structure} of fluctuations on \emph{large linear scales}, in a way that improves over the Wiener filter alone. The morphology of line-intensity fluctuations on large scales, separate from their amplitude, is still of interest for many of the reasons we discussed in the~\hyperref[sec:intro]{Introduction}.

Of course, we would like to show that the LCB filter improves over the Wiener filter in recovery of $P(k)$ amplitudes as well. However, to quantify the success of a given signal estimator $\hat{s}$ in recovering structure regardless of amplitude, we will calculate the normalised cross-correlation between the estimator and the original signal, as a function of $k$. Specifically, if the original signal has a power spectrum $P^{(S)}(k)$, the reconstruction has a power spectrum $P^{(\hat{s})}(k)$, and the signal--reconstruction cross power spectrum is $P^{(S\times\hat{s})}(k)$, the normalised cross-correlation is
\begin{equation}
    r(k)\equiv\frac{P^{(S\times\hat{s})}(k)}{\sqrt{P^{(S)}(k)P^{(\hat{s})}(k)}}.
\end{equation}
Perfect recovery of the phases of the Fourier modes, regardless of amplitude, corresponds to $r=1$.

\section{Simulations of \texorpdfstring{[C\textsc{\,II}]}{CII} LIM reconstruction---proof of concept}
\label{sec:CIILIM}
Having considered the formalism around the LCB filter and how to quantitatively establish successful reconstruction, we are now ready to consider applications to LIM reconstruction. Our first simulated case study is a futuristic [C\textsc{\,ii}] LIM survey that has a similarly futuristic catalogue of $z=5.5$--6.6 Lyman-alpha emitters (LAE) available for cross-correlation. We will begin by explaining the scientific context behind [C\textsc{\,ii}] LIM in~\autoref{sec:cii_context}, before defining the survey model in~\autoref{sec:CIInoise} and the line emission signal and interloper models in~\autoref{sec:ciimodels}. We will then be prepared to consider reconstruction results in simulations in~\autoref{sec:cii_results}.

For this section, we assume cosmological parameter values of $\Omega_m = 0.307$, $\Omega_\Lambda = 0.693$, $\Omega_b =0.048$, $h=0.678$, $\sigma_8 =0.829$, and $n_s =0.96$, to maintain consistency with the Small MultiDark-Planck (SMDPL) simulation~\citep{MultiDark} that we will use below. This cosmology is also consistent with the Planck 2015 results~\citep{Planck15}.

\subsection{Context---why ionised carbon and Lyman-alpha emitters?}
\label{sec:cii_context}
The [C\textsc{\,ii}] line, emitting at a rest-frame frequency of 1900.5 GHz, is of interest at these high redshifts as a potential signpost for star formation and diffuse interstellar gas in the late epoch of reionisation. [C\textsc{\,ii}] is a key cooling line for the interstellar media of galaxies, accounting for as much as 1\% of the total far-infrared luminosity~\citep{Stacey91}, and correlates with star-formation rate in population studies, both in nearby galaxies~\citep{DL14,HC15} and at redshifts reaching into the tail end of the epoch of reionisation~\citep{Schaerer20}. [C\textsc{\,ii}] also traces ionised and diffuse neutral phases of the interstellar medium, as shown by detailed line diagnostics in nearby galaxies~\citep{Croxall17} and extended emission at high redshift~\citep{Carniani18,Fujimoto19,Matthee19,Rybak19,Meyer22}, which has motivated use of [C\textsc{\,ii}] as a tracer of neutral or H\textsc{\,i} gas in galaxies across cosmic history~\citep{Heintz21,Heintz22}. Whether it traces the assembly of the first star-forming galaxies assembling post-reionisation or neutral gas accreting into galaxies, the [C\textsc{\,ii}] LIM signal has strong complementarity with 21 cm tomography and provides a key probe of the late epoch of reionisation.

Searches for [C\textsc{\,ii}] emission from $z\simeq2.5$ using cross-correlation of Planck 545 GHz data with quasar positions have yielded tentative results~\citep{Pullen18,Yang19} but depend heavily on accurate models of correlated continuum emission due to the broadband nature of the intensity data. This is a problem not shared to the same extent by LIM surveys with many ($\gtrsim10^2$) channels, which should confine correlated continuum emission to the lowest-$k$ line-of-sight modes and thus be theoretically capable of rejecting correlated continuum with minimal signal loss~\citep{Switzer19}. (We will neglect correlated continuum contamination for the simulations in the remainder of this work, which for our purposes should be equivalent to assuming that the channel count of these surveys allow appropriate de-biasing of signal power spectra to correctly scale the reconstructed line intensity.) This has motivated the development of numerous [C\textsc{\,ii}] LIM pathfinder experiments~\citep{CONCERTO,Crites14,CCATp}.

However, LIM observations of [C\textsc{\,ii}] from $z\sim6$ (so at $\simeq270$ GHz) will see significant interloper line emission from the rotational transitions of CO at lower redshift. The CO($J\to J-1$) lines, which have rest-frame frequencies of $\simeq115\cdot J$ GHz, trace cold molecular gas that fuels star formation and are thus interesting in their own right, and we will discuss CO LIM in~\autoref{sec:COLIM}. However, the interloper CO(3--2) through CO(7--6) emission components are a significant concern for [C\textsc{\,ii}] surveys as they originate from redshifts 0.3--2.0, parts of cosmic history that are full of much closer, more metal-rich, comparatively more massive, and more actively star-forming galaxies compared to $z\sim6$. We may also consider the neutral carbon [C\textsc{\,i}](1--0) line (rest-frame frequency of 492.16 GHz) as an interloper along the same lines as the CO lines, as it too acts as a tracer of cold molecular gas~\cite{Jiao17,Valentino18,Nesvadba19}. Previous literature has extensively explored different ways of reducing interloper bias in the measured [C\textsc{\,ii}] power spectrum~\citep{Breysse15,LidzTaylor16,Cheng16,Sun18,Cheng20}, but here we consider the even more ambitious possibility of reconstructing the [C\textsc{\,ii}] fluctuations themselves without interloper bias by leveraging cross-correlation against a hypothetical LAE catalogue.

We choose to simulate a LAE catalogue because previous surveys (e.g., Refs.~\cite{MUSEWide,SILVERRUSH}) suggest LAEs are abundant at these redshifts, with~\textcite{SILVERRUSH} for example finding $>50$ LAEs per square degree per $\Delta z\simeq0.1$. Combining better photometry with next-generation IFUs or massively multiplexed spectrographs, future surveys could identify thousands of LAEs across wide fields and wide continuous redshift ranges.

However, note that for the purposes of this work, what matters most is not the exact galaxy selection for the most part, but rather the source density, the correlation of source density contrast with the signal to be reconstructed, and the effective linear bias of the galaxy sample. This last factor affects the amplitude of the cross power spectrum, thus affecting detectability for fixed survey sensitivity. Ultimately, even if a wide-field flux-limited LAE survey is not feasible, any spectroscopic survey with similar source abundances that selects for similarly star-forming galaxies would serve equally well as a cross-correlation target for some future LIM survey.

\subsection{Survey parameters and Gaussian noise model}
\label{sec:CIInoise}
The futuristic [C\textsc{\,ii}] survey that we simulate here takes its cues from the proposed roadmap of~\textcite{Karkare22}, which suggests that future stages of [C\textsc{\,ii}] surveys should increase geometrically in sensitivity. The futuristic survey concept considered in Ref.~\cite{selfcite_lims3} fits in Stage 3 of the roadmap of~\textcite{Karkare22}, with $N_\text{spec}t_\text{surv}=1.8\times10^8$ spectrometer-hours covering 1024 square degrees of sky.

We follow Ref.~\cite{selfcite_lims3} and calculate the noise-equivalent intensity per detector as
\begin{equation}
    \sigma_\text{spec}=10^6\text{ Jy\,sr}^{-1}\,\text{s}^{1/2}\left(\frac{\delta\nu}{2.5\,\text{GHz}}\right)^{-1/2}\left(\frac{\epsilon_\text{sys}}{0.05}\right),
\end{equation} 
given channel bandwidth $\delta\nu$ and system emissivity $\epsilon_\text{sys}$, and under the implicit assumption that the detection bandwidth is equal to the frequency channel width. Unlike Ref.~\cite{selfcite_lims3}, we will assume $\epsilon_\text{sys}=0.05$ for simplicity, and we will assume $\delta\nu=312.5$ MHz for added futurism.

Assuming adequate control of atmospheric and other sources of frequency-dependent Earth-bound noise, the final `thermal' map noise per voxel should then be
\begin{equation}
    \sigma_N = \frac{\sigma_\text{spec}}{\sqrt{N_\text{spec}t_\text{surv}\Omega_\text{pix}/\Omega_\text{surv}}}.
\end{equation}
While we assume the LIM survey covers a total of 1024 square degrees of sky, we conservatively assume only a square 9 deg$^2$ subfield can be cross-correlated with an overlapping LAE catalogue and thus reconstructed with the LCB filter. This $3\times3$ deg$^2$ patch will be simulated with $512\times512$ pixels, for a solid angle per pixel of $\Omega_\text{pix}\simeq1.05\times10^{-8}$ sr. With all appropriate substitutions made, we find $\sigma_N\simeq1.9\times10^4$ Jy sr$^{-1}$.

We also assume a Gaussian beam profile with a full width at half maximum (FWHM) of 50 arcseconds, taken to be achromatic for simplicity. This approximately corresponds to the diffraction limit for a 6 m telescope observing at 250 GHz, so it is quite plausible to intentionally degrade the observation to that common resolution across all frequencies after the fact (e.g., to avoid mode mixing where possible).
\subsection{Simulation parameters and line emission models}
\label{sec:ciimodels}
For [C\textsc{\,ii}] simulations, our starting point is the SMDPL simulation, which is a cosmological $N$-body box of comoving size $400h^{-1}$ Mpc and mass resolution $9.6\times10^7h^{-1}\,M_\odot$. More precisely, the starting point relevant to this work is the catalogue of dark matter halos identified in the simulation across space and time. Given the mass resolution, the halo population should be reasonably complete down to virial masses of $M_h\simeq10^{10}\,M_\odot$; we set this as the minimum halo mass for non-zero line luminosity.

We use publicly available halo catalogues (online at time of writing at \url{https://www.peterbehroozi.com/data.html}) that have already been processed through the UniverseMachine framework of~\textcite{UM}. This framework provides an empirical galaxy model based on associating halo merger histories with star formation, and as a result assigns galaxy properties such as star-formation rate (SFR) and dust extinction ($A_\text{UV}$) to every halo in each simulation snapshot. The UniverseMachine release includes code to generate lightcones from these halo catalogues, so we generate 10 lightcones spanning $3\times3$ deg$^2$ and $z=0$--9. These will be useful for all kinds of LIM predictions, especially given the mass resolution of SMDPL, but we leave further exploration to future work.

We convert the galaxy properties generated by UniverseMachine into line luminosities relevant to our [C\textsc{\,ii}] survey in the frequency range of 250--290 GHz, which include not only the [C\textsc{\,ii}] line from redshifts $z=5.5$--6.6, but also CO rotational transitions as well as the atomic carbon [C\textsc{\,i}] line. We use correlations between SFR and line luminosities summarised by Ref.~\cite{selfcite_lims3}, which in turn adapts all of the relevant relations from the work of~\textcite{Sun21}. We briefly resummarise them here:
\begin{itemize}
    \item The [C\textsc{\,ii}] luminosity is assumed on average to be linearly proportional to the SFR (via UV luminosity), with the proportionality derived from combining values given by Refs.~\cite{Sun21,SunFurlanetto16}:
    \begin{equation}
        \log\left(\frac{L_\text{[C\textsc{\,ii}]}}{L_\odot}\right) =  \log{\left(\frac{\mathrm{SFR}}{M_\odot\text{ yr}^{-1}}\right)}+7.34.
    \end{equation}
    In addition, we assume a log-normal scatter around this average with a standard deviation of $0.4$ (in units of dex).
    \item The CO transitions and the [C\textsc{\,i}] line are all assumed to correlate perfectly with the CO(1--0) luminosity, which in turn is assumed on average to be related to the SFR (via IR luminosity) through a power law:
\begin{align}
    \log\left(\frac{L_{\text{CO(1--0)}}}{L_\odot}\right) &= \alpha_\text{IR--CO}^{-1}\left[\log{\left(\frac{\mathrm{SFR}}{M_\odot\text{ yr}^{-1}}\right)}\right.\nonumber\\&\quad\left.{}+9.76-\beta_\text{IR--CO}\right]-4.31,
\end{align}
    where $\alpha_\text{IR--CO}=1.27$ and $\beta_\text{IR--CO}=-1.0$. We also assume a log-normal scatter around this average with a standard deviation of $0.4$ (again in units of dex). From this scattered CO(1--0) luminosity, the luminosities for the CO($J\to J-1$) lines can be calculated as follows:
\begin{align}
    L_{\text{CO}(J\to J-1)} = r_JJ^3L_{\text{CO(1--0)}},
\end{align}
where we use the values of $r_3=0.73$, $r_4=0.57$, $r_5=0.32$, $r_6=0.19$, and $r_7=0.1$. We essentially calculate the [C\textsc{\,i}] luminosity in the same way, treating it as if it were another CO line with $J_\text{[C\textsc{\,i}]}\equiv\nu_\text{rest,[C\textsc{\,i}]}/\nu_\text{rest,{CO(1--0)}}=4.27$:\begin{align}
    L_{\text{[C\textsc{\,i}]}} = r_\text{[C\textsc{\,i}]}J_\text{[C\textsc{\,i}]}^3L_{\text{CO(1--0)}},
\end{align}
where $r_\text{[C\textsc{\,i}]}=0.18$.

All line ratios relative to the CO(1--0) luminosity are taken from~\textcite{Sun21}, who in turn refer the reader to Refs.~\cite{Kamenetzky16,Valentino18,Nesvadba19} for observational literature probing the correlations of the individual line species with IR luminosity.
\end{itemize}

In addition, we also need to estimate Lyman-alpha luminosities in order to simulate a flux-limited (and therefore luminosity-limited) LAE sample. We once again look to~\textcite{Sun21}, who provide an approximate relation between the SFR and the Lyman-alpha luminosity assuming equilibrium between ionisation and recombination, with the Lyman-alpha photons coming from recombination events only. Given $f_\gamma=4000$ ionising photons produced per stellar baryon, an escape fraction $f_\text{esc}=0.1$ of ionising photons that thus do not lead to recombination events, a fraction $f_{\text{Ly}\alpha}=0.67$ of recombination events producing Lyman-alpha emission, and an escape fraction $f_\text{esc}^{\text{Ly}\alpha}=0.6$ of Lyman-alpha photons that reach the observer,
\begin{equation}
    L_{\text{Ly}\alpha}=\frac{f_\gamma\,\mathrm{SFR}/(10^{A_\text{UV}/2.5})}{m_p/(1-Y)}(1-f_\text{esc})f_{\text{Ly}\alpha}f_\text{esc}^{\text{Ly}\alpha}E_{\text{Ly}\alpha}.
\end{equation}
Here, $m_p$ is the proton mass, $Y=0.24$ is the helium mass fraction, $A_\text{UV}$ is the dust extinction (recalling that UniverseMachine calculates $A_\text{UV}$ alongside SFR for each halo), and $E_{\text{Ly}\alpha}=hc/(1215\,\text{\AA})$ is the energy per Lyman-alpha photon. As~\textcite{Sun21} acknowledge, the resulting model still ignores environmental factors that affect the distribution of LAEs, but matches luminosity functions and tracer bias constraints in the literature derived from a 14--21 deg$^2$ Subaru Hyper-Suprime Cam (HSC) survey~\citep{SILVERRUSH}. This is sufficient for our purposes as we mostly want to reproduce realistic source densities and biases rather than exact source selections, at least in the context of the present work.

We use \texttt{limlam\_mocker} (available at time of writing at \url{https://github.com/georgestein/limlam\_mocker}) to load each simulation lightcone, generate line luminosities for each halo, and bin the [C\textsc{\,ii}] and interloper line luminosities into cubes of flux density (in units of Jy sr$^{-1}$) based on the redshift-space locations of the halos. The LIM survey instrument is assumed to have a Gaussian beam on the sky with 50 arcsecond FWHM, so we convolve the flux density cube with the appropriate Gaussian kernel. We also subtract the mean from all flux density cubes, as LIM surveys tend to remove the mean level in analysis while subtracting continuum foregrounds. We then have a pseudo-intensity cube with 128 frequency channels spanning 250--290 GHz and $512$ pixels along each angular dimension, to which we then add Gaussian noise with standard deviation $\sigma_N$ derived in~\autoref{sec:CIInoise}.

At the same time, we generate a mock LAE catalogue corresponding to each flux density cube by selecting only halos with an associated Lyman-alpha luminosity of $L_{\text{Ly}\alpha}>10^{42.5}$ erg s$^{-1}$. This is somewhat more conservative than the cut of $10^{42.2}$--$10^{42.3}$ erg s$^{-1}$ used by~\textcite{Sun21} for their mock LAE survey. Note however that we are hoping for a wide-field survey to generate this LAE catalogue across the whole redshift range, which likely will not be possible with the same source density as the Subaru HSC narrow-band survey of Refs.~\cite{SILVERRUSH,SILVERRUSHLF} (which only spans two redshift bands of $\Delta z\simeq0.1$ each). The resulting catalogues provide between 7700 and 8600 LAE positions across our 10 lightcones.

We bin the LAEs into a density contrast cube, obtaining the source count in each voxel but then dividing by the mean and subtracting 1. We may now obtain total auto and cross spectra with this LAE density contrast cube and the LIM data cube, for each one of our 10 semi-independent realisations.

The LCB filter, like the Wiener filter, requires knowledge of the signal power spectrum. We obtain a form for the pseudo-intensity power spectrum (i.e., the power spectrum of the [C\textsc{\,ii}] flux density field after convolution with the instrument beam and mean subtraction) by averaging across all 10 lightcones, basically assuming that there is a reasonable estimate of at least the shape of the [C\textsc{\,ii}] power spectrum from the wider LIM survey.

\subsection{Simulation results}
\label{sec:cii_results}

\begin{figure*}[t!]
    \centering
    \includegraphics[width=0.96\linewidth]{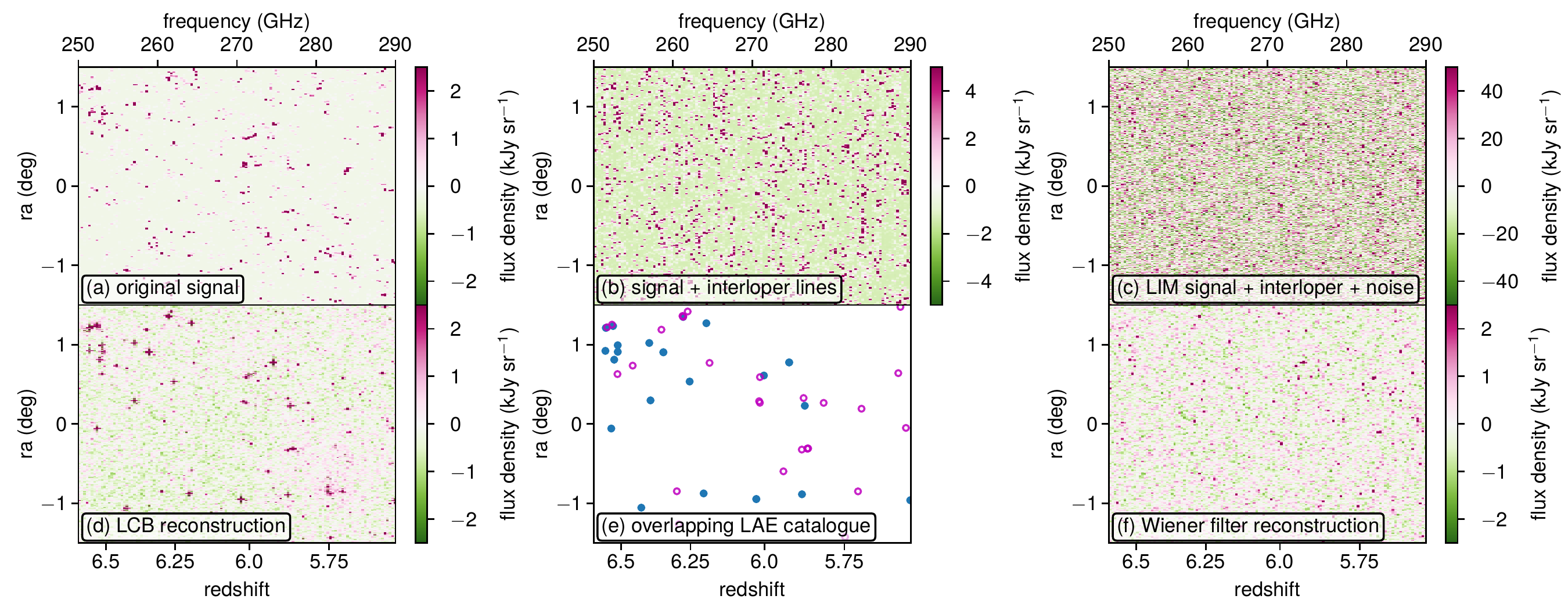}
    \caption{Visualisation of a slice of an example [C\textsc{\,ii}] LIM simulation, showing six outputs: (a) the mean-subtracted [C\textsc{\,ii}] signal in isolation; (b) the mean-subtracted [C\textsc{\,ii}] signal summed with the mean-subtracted interloper CO and [C\textsc{\,i}] emission; (c) the total mean-subtracted signal and interloper emission plus Gaussian noise as described in~\cref{sec:CIInoise}; (d) the LCB-filtered data; (e) the LAE positions contained in the slice shown of the data cube, including both sources properly contained in the slice's declination bin (filled circles) and sources in neighbouring declination bins (unfilled circles); and (f) the Wiener-filtered data. Both the LCB and Wiener filter reconstructions assume full knowledge of the signal power spectrum.}
    \label{fig:CIIexample}
\end{figure*}
Before considering the filtering results, for completeness we note the signal-to-noise ratio summed across all $k$ for both the [C\textsc{\,ii}] auto and [C\textsc{\,ii}]--LAE cross power spectra (assuming successful removal of interloper line emission in the former case). If $N_m$ modes are available to the observation at wavenumber $k$, the variance for each power spectrum is given by
\begin{align}
    \sigma^2[P^{(S)}]&= \frac{\left(P^{(S)}+P^{(N)}\right)^2}{N_m};\\\sigma^2[P_{12}^{(S)}]&= \frac{\left(P_1^{(S)}+P_2^{(N)}\right)\left(P^{(S)}+P^{(N)}\right)+{P_{12}^{(S)}}^2}{2N_m}.
\end{align}
with $d_1$ and $d_2$ respectively being the LAE and LIM observations. Note that in this case $P^{(N)}$ comes from the sum of the thermal noise described in~\autoref{sec:CIInoise} and the interloper line emission.

The total signal-to-noise ratio for a given power spectrum $P(k)$ is then given by the signal-to-noise ratio in each $k$-bin summed in quadrature:
\begin{equation}
    (\mathrm{S/N})[P(k)] = \left(\sum_k\frac{P^2(k)}{\sigma^2[P(k)]}\right)^{1/2}.
\end{equation}
Across our 10 different realisations, this total signal-to-noise ratio ranges from 91 to 117 for the [C\textsc{\,ii}] auto power spectrum in the 9 deg$^2$ field alone (recalling that we have imagined a survey that spans over 1000 deg$^2$ in total), and from 312 to 337 for the [C\textsc{\,ii}]--LAE cross power spectrum. While detectability forecasting is not the focus of the present work, this calculation shows that there is significant information content in the cross power spectrum of our mock surveys that the LCB filter will use but the Wiener filter will simply ignore.

\subsubsection{Filtering with full knowledge of signal power spectrum}

To start with, we assume that the power spectrum $P^{(S)}$ of the signal is fully known---possibly from the wider LIM survey---and used to design the LCB filter, as well as a Wiener filter for comparison. We first find the average signal power spectrum across all 10 lightcones, then use it while iterating through the lightcones again to design the filters in each simulation, along with the actual observed total auto and cross power spectra from that simulation. We show an example set of observables and reconstructions under this implementation in~\cref{fig:CIIexample}.

Note first the strength of the total interloper line intensity contrast---which significantly exceeds that of the [C\textsc{\,ii}] intensity contrast. The Wiener filter, lacking the information needed to distinguish between the signal and interloper lines, ends up reconstructing neither component. The LCB filter, however, is able to leverage the cross-correlation between the signal component and the LAE catalogue to isolate certain bits of flux density contrast that likely originate from $z\sim6$ [C\textsc{\,ii}]. The resulting reconstruction is not perfect, with some spurious bright spots and failure to recover bright spots without a co-located LAE overdensity. However, the large-scale fluctuations certainly appear soundly isolated from the interloper emission as well as the noise.

\begin{figure}
    \centering
    \includegraphics[width=0.96\linewidth]{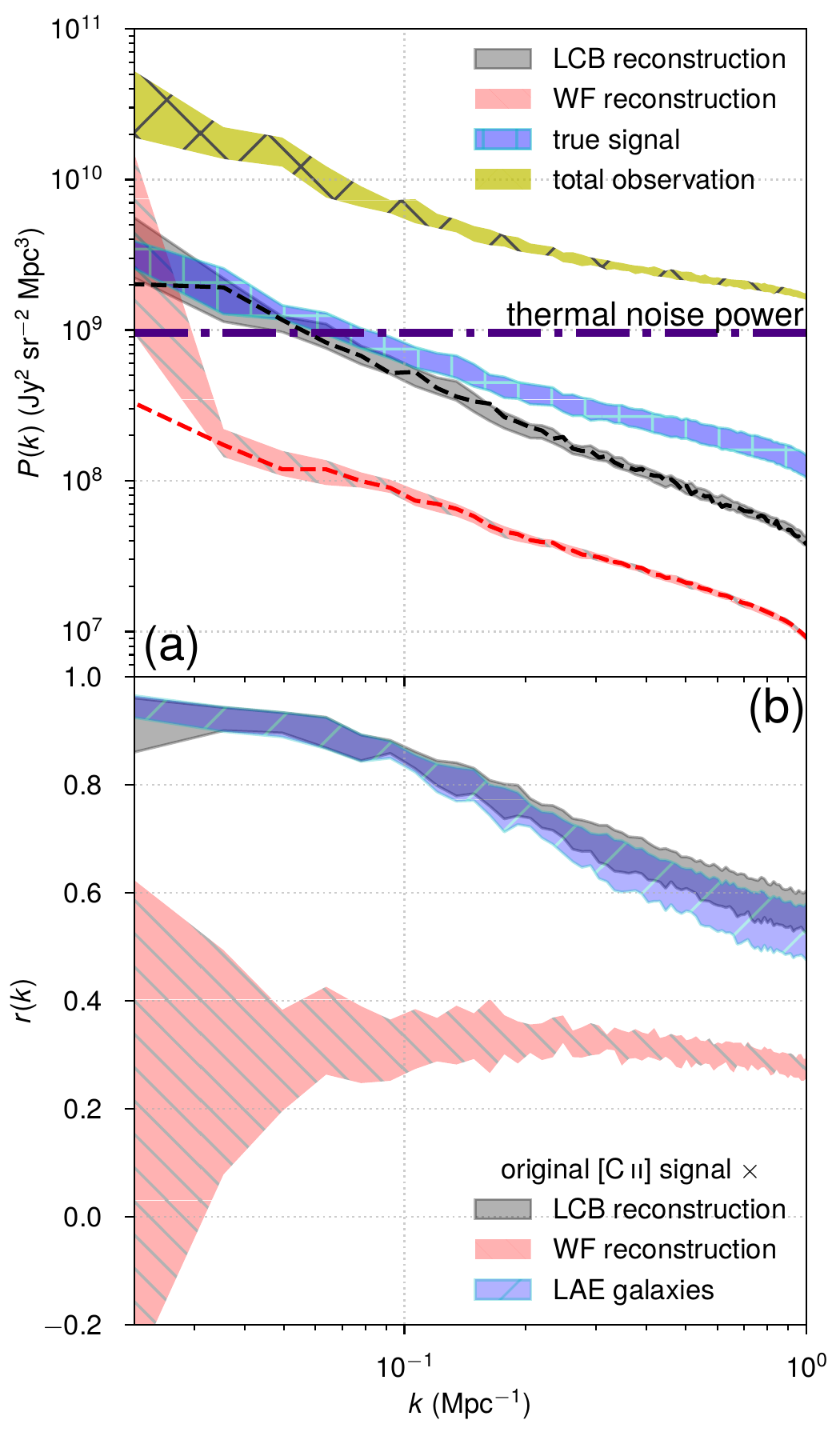}
    \caption{The full interval of [C\textsc{\,ii}] reconstruction results across our 10 semi-independent lightcones. We show (a) the obtained power spectra of the LCB and Wiener filter reconstructions alongside their expected values (dashed lines), as well as the original input signal, the total observation including noise and interloper emission, and the thermal noise power spectrum by itself (dash-dotted line). We also show (b) the normalised cross-correlation $r(k)$ between the original input signal and the two reconstructions, as well as between the [C\textsc{\,ii}] emission and the LAE density contrast.}
    \label{fig:Pk_CII}
\end{figure}

Moving from first impressions to quantitative results, we show in~\cref{fig:Pk_CII} the range across all 10 simulations of relevant power spectra and normalised cross-correlation. Before considering the reconstructions, it is worth noting that in this futuristic LIM survey, the limiting factor in sensitivity to the [C\textsc{\,ii}] power spectrum is not the thermal Gaussian noise, but the interloper line emission (which is what accounts for the difference between the total observed power spectrum and the combination of the true signal and thermal noise power spectra). In the presence of thermal noise alone, the Wiener filter would likely actually succeed in reconstructing at least the largest scales. However, with the interloper emission power spectrum being ten times that of the signal power spectrum, the Wiener filter reconstruction is inevitably suppressed in amplitude by the corresponding amount. The normalised cross-correlation with the true signal is also poor, mostly staying near 30--40\%.

Using the measured [C\textsc{\,ii}]--LAE cross-correlation, the LCB filter clearly improves reconstruction of the [C\textsc{\,ii}] LIM signal. Compared to the Wiener filter, the LCB filter greatly reduces the power spectrum bias. More importantly, in the range of $k=0.02$--1 Mpc$^{-1}$ (between the lowest $k$ available given the survey volume and the highest meaningful $k$ given the survey resolution), the normalised cross-correlation between the LCB filter reconstruction and the true signal never falls below 50\%, and reaches 80--90\% at the linear scales of $k\lesssim0.1$ Mpc$^{-1}$.

Unsurprisingly, the normalised cross-correlation between the LCB reconstruction and the true signal is very similar to that between the LAE catalogue and the true signal. This matches our first impressions from visual inspection of~\cref{fig:CIIexample}, in that we largely isolate the portion of the [C\textsc{\,II}] signal that correlates with the LAE galaxies. Knowledge of the expected [C\textsc{\,ii}] power spectrum still aids the LCB filter, but does not contribute as much to the reconstruction as the cross-correlation, at least in this case.

\subsubsection{Filtering with knowledge of shape but not amplitude of signal power spectrum}

Now we consider a more realistic scenario where the power spectrum $P^{(S)}$ is not in fact fully known. In principle, if [C\textsc{\,ii}] traces matter density fluctuations, the shape of the [C\textsc{\,ii}] power spectrum could be known beforehand to a great extent. The amplitude, however, could be imperfectly determined. The presence of interloper bias and other sources of additive noise could result in overestimation, but over-correcting for these sources of biases could result in underestimation.

\begin{figure}
    \centering
    \includegraphics[width=0.96\linewidth]{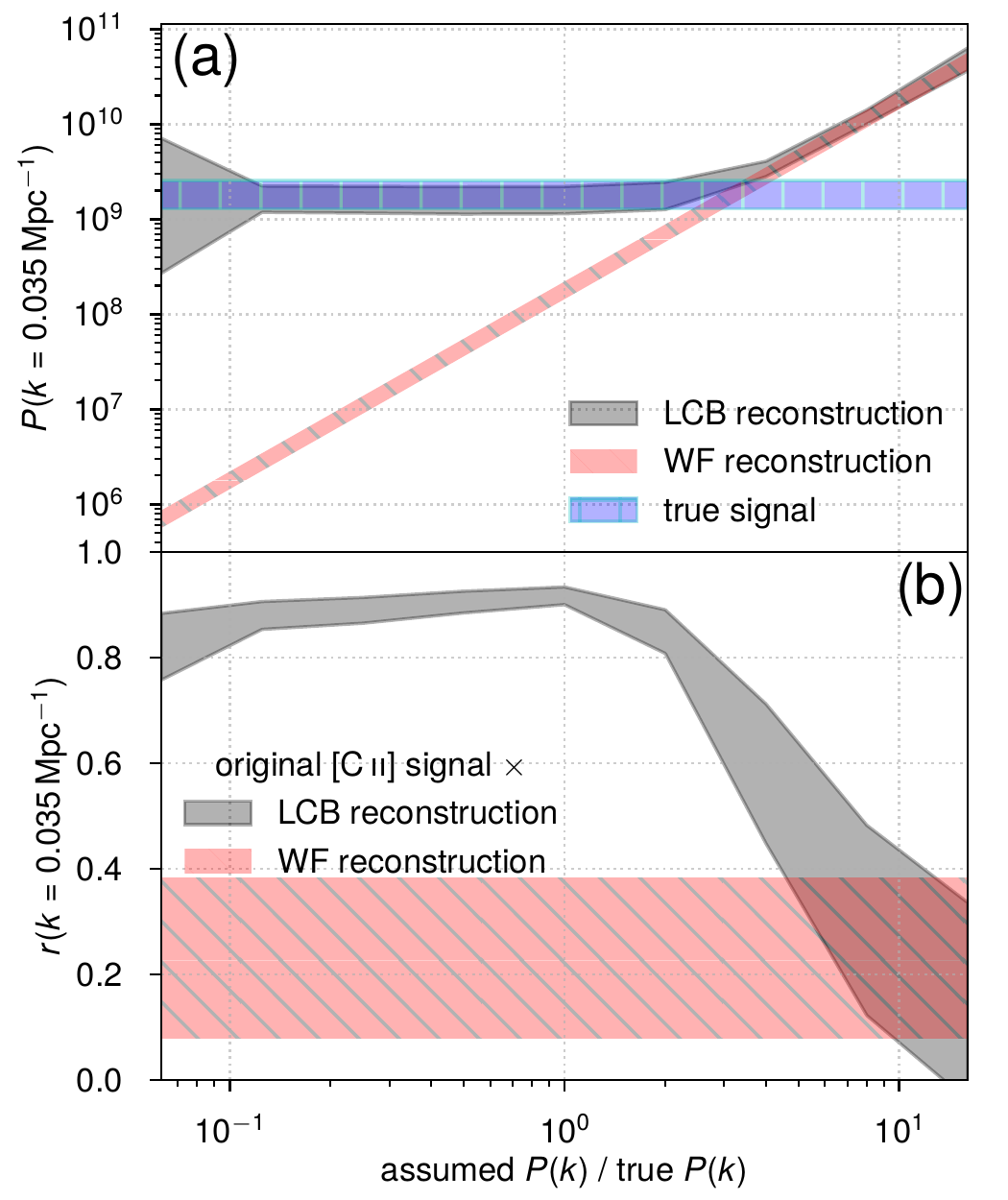}
    \caption{Effect of calculating LCB and Wiener filter reconstructions of the [C\textsc{\,ii}] signal with an assumed power spectrum $P^{(S)}(k)$ for the signal that differs from the true $P^{(S)}(k)$ by some multiplicative offset indicated along the $x$-axis. As a function of this offset, we show at $k=0.035$ Mpc$^{-1}$ the range across our 10 simulated observations of (a) the value of the power spectra of the LCB and WF reconstructions (with the true signal power plotted for comparison), and (b) the normalised cross-correlation of each reconstruction with the true [C\textsc{\,ii}] signal.}
    \label{fig:Pkbias_CII_lowk}
\end{figure}

We thus consider calculating LCB and Wiener filter coefficients based on an assumed $P^{(S)}(k)$ that is off by some multiplicative factor from the true $P^{(S)}(k)$, and examining the resulting LCB and WF reconstructions. We show the resulting change in the reconstruction at $k=0.035$\,Mpc$^{-1}$ in~\cref{fig:Pkbias_CII_lowk}, with the multiplicative offset ranging from $1/16$ to 16.

Neither filter is robust against overestimation of $P^{(S)}(k)$. This is sensible given the definitions of the filter coefficients, as noisy data will not be sufficiently down-weighted if we overestimate the contribution of the target signal to the total measured power spectrum.

However, the LCB filter \emph{is} robust against underestimation of $P^{(S)}(k)$. Whereas the Wiener filter would na\"{i}vely down-weight what is in fact perfectly serviceable data, the LCB filter is anchored by the presence of external data and cross-correlation of the signal against those data. Therefore, even if we lose trust in the [C\textsc{\,ii}] auto power spectrum, the [C\textsc{\,ii}]--galaxy cross spectrum and the galaxy auto spectrum together limit the resulting bias in the reconstruction.

This distinction is visible not only in the power spectrum of the LCB and WF reconstructions, but in their normalised cross-correlation against the true signal. Regardless of the exact coefficients, the Wiener filter is simply re-weighting the same data, and so there is no reason for $r(k)$ to change with the filter coefficients. More generally, this suggests that in the presence of strong interloper emission, a Wiener filter will never reconstruct the Fourier modes of the [C\textsc{\,ii}] fluctuations with high fidelity.

\begin{figure}
    \centering
    \includegraphics[width=0.96\linewidth]{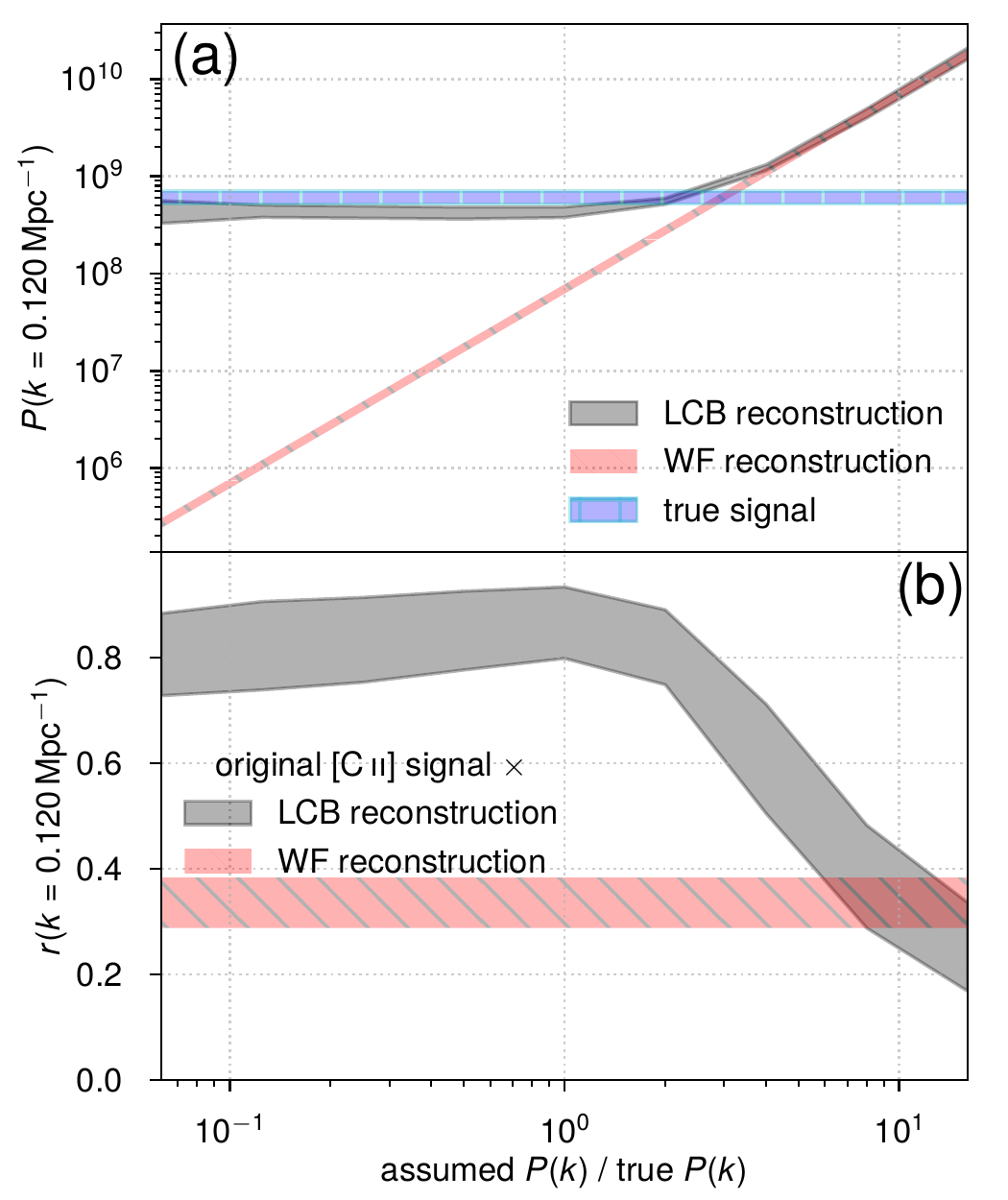}
    \caption{Same as~\cref{fig:Pkbias_CII_lowk}, showing evolution with signal power spectrum mis-estimation of (a) power spectra of reconstructions (with the true signal power plotted for comparison), and (b) normalised cross-correlation, but all at a higher wavenumber of $k=0.120$ Mpc$^{-1}$.}
    \label{fig:Pkbias_CII_slightlylesslowk}
\end{figure}

However, while overestimation of the power spectrum by beyond a factor of two rapidly degrades the LCB reconstruction as well, underestimation by as much as a factor of eight has relatively little effect, with $r(k=0.035$\,Mpc$^{-1})$ against the true signal staying above 80\% thanks to the LAE cross-correlation filling the vacuum left by the down-weighted LIM data. We show in~\cref{fig:Pkbias_CII_slightlylesslowk} the same results at a slightly higher $k=0.120$\,Mpc$^{-1}$ to show some degree of $k$-independence in these qualitative conclusions about the merits of the LCB filter.

Given its robust performance, the LCB filter would likely equally excel at recovering the interloper emission components individually, but we leave exploration of this possibility to future work.

\section{Potential applications in CO LIM}
\label{sec:COLIM}
Different LIM experiments have different sets of foregrounds and interloper emission components, and here we will consider different possible future configurations of the CO Mapping Array Project (COMAP) as the focus of our remaining simulations. We provide the context behind COMAP first in~\autoref{sec:COMAP}, then consider potential use cases for the LCB filter in near-term and future phases of COMAP operations respectively in~\autoref{sec:COMAPPF} and~\autoref{sec:COMAPER}.

For this section, we assume cosmological parameter values of $\Omega_m = 0.286$, $\Omega_\Lambda = 0.714$, $\Omega_b =0.047$, $h=0.7$, $\sigma_8 =0.82$, and $n_s =0.96$, to maintain consistency with previous simulations used by Ref.~\cite{Ihle19}.

\subsection{Context---the COMAP Pathfinder-like survey, the COMAP-ERA concept, and a fiducial CO model}
\label{sec:COMAP}

\subsubsection{COMAP roadmap and survey parameters}

As we noted in~\autoref{sec:cii_context}, the rotational transition lines of CO trace the cold molecular gas that fuels star formation~\cite{CW13}. Therefore, CO LIM has the potential to trace the evolution of the early star-forming environments from the late epoch of reionisation to the epoch of peak star formation and galaxy assembly.

COMAP is currently the only operating LIM experiment with dedicated single-dish instrumentation targeting the low-$J$ CO lines, having deployed a Pathfinder experiment in 2019 and published early science results from the first observing season as summarised by Ref.~\cite{Cleary22}. While the current Pathfinder phase of COMAP operates in the Ka band and covers 26--34 GHz in observing frequency, this corresponds to not only CO(1--0) emission from $z=2.4$--3.4 (the main science target of the COMAP Pathfinder), but also to CO(2--1) emission from $z=5.8$--7.9. Future phases of COMAP as described in Ref.~\cite{Breysse22} will deploy Ku-band instruments observing at 12--20 GHz, thus measuring CO(1--0) at $z=4.8$--8.6 and allowing cross-correlation between Ka- and Ku-band observations to better constrain cosmic molecular gas content at $z\sim7$.

\begin{table*}
    \centering
    \caption{Key parameters for COMAP Pathfinder-like and COMAP-ERA simulations, either taken from~\protect\cite{Breysse22} or newly assumed for this work.}
    \label{tab:comapparams}
    \begin{ruledtabular}
    \begin{tabular}{llccc}
          & & COMAP Pathfinder-like & \multicolumn{2}{c}{COMAP-ERA:} \\\cline{4-5}
         Parameter &  & Ka band & Ku band & Ka band\\\hline
         Frequency coverage & (GHz) & 26--34 &  13--17 & 26--34\\
         Beam FWHM & (arcmin) & 4.5 & 3.9 & 4.5\\
         Pixel size $\Omega_\text{pix}$ & (arcmin$^2$) & $4\times4$ & $4\times4$ & $4\times4$\\
         Science channelisation $\delta\nu$ & (MHz) & 15.625 & 7.8125 & 15.625\\
         Nominal system temperature $T_\text{sys}$ & (K) & 44 & 20 & 44\\
         Spectrometer count per dish $N_\text{spd}$ & & 19 & 38 & 19\\
         Solid angle per field $\Omega_\text{obs}$ & (deg$^2$) & $2\times2$ & $2\times2$ & $2\times2$\\
         Dish-hours per field $N_\text{dish}t_\text{obs}$ & (hr) & 12500 & 57000 & 110000\\
         Noise per voxel $\sigma_N$ & (\textmu K) & 11.42 & 2.55 & 3.85\\
         Number of fields per survey & & 3 & 3 & 3
    \end{tabular}
    \end{ruledtabular}
\end{table*}

\autoref{tab:comapparams} describes the parameters we assume for two phases of COMAP.
\begin{itemize}
    \item The first is a COMAP Pathfinder-like survey, with slightly increased integration time per field compared to the fiducial Pathfinder survey design. This sensitivity could be readily achieved with extended Pathfinder observations and the addition of a second instrument would further facilitate matters. 
    \item The second is the COMAP Expanded Reionisation Array (COMAP-ERA), a hypothetical Stage 2 experiment (if we index the Pathfinder as Stage 0) with most of the survey parameters as outlined in Ref.~\cite{Breysse22}.
\end{itemize}
In both cases, we alter the voxel size so that the pixel size is roughly the beam size and the channel width significantly exceeds the native $\simeq2$ MHz resolution of the COMAP spectrometers. Also in both cases, the noise per voxel is calculated as
\begin{equation}
    \sigma_N = \frac{T_\text{sys}}{\sqrt{\delta\nu\,N_\text{spd}N_\text{dish}t_\text{obs}\Omega_\text{pix}/\Omega_\text{obs}}}.
\end{equation}

\subsubsection{CO model and simulation parameters}
\label{sec:CO_sims}
We must still define a model for CO emission and apply it to an ensemble of simulations. Unlike in~\cref{sec:CIILIM}, where we used UniverseMachine lightcones from a single cosmological simulation with empirically determined star-formation rates for each halo, here we will make use of an ensemble of independent approximate $N$-body simulations using the peak-patch method~\citep{mPP}. Without merger histories for peak-patch halos, we will simply associate the halo mass $M_h$ and cosmological redshift $z$ to an average star-formation rate and the star-formation rate to an average CO line luminosity.

For the CO luminosity model, we use the model of~\textcite{Li16} as modified by~\textcite{mmIME-ACA}, assigning luminosities to all halos with virial mass above $10^{10}\,M_\odot$.
\begin{itemize}
    \item The model begins by using the empirical model of~\textcite{Behroozi13a,Behroozi13b} to interpolate the average star-formation rate for a given halo mass and redshift. Each halo is then assigned its appropriate star-formation rate with log-normal scatter of $\sigma=0.3$ dex (although we implement the scatter in a way that preserves the linear mean).
    \item The star-formation rate is assumed to be linearly proportional to the bolometric IR luminosity, via a factor of $10^{10}\,L_\odot\,(M_\odot$ yr$^{-1})^{-1}$.
    \item Empirical fits to observations relate the IR and average CO($J\to J-1$) luminosities:
    \begin{equation}
    \log{\frac{L_\text{IR}}{L_\odot}} = \alpha_J\left(\log{\frac{L_{\text{CO},J}}{L_\odot}}+4.31\right)+\beta_J.
    \end{equation}
    Specifically, following fits to a local sample of galaxies~\textcite{Kamenetzky16}, we use $\alpha_1=1.27$, $\beta_1=-1.0$, $\alpha_2=1.11$, and $\beta_2=-0.6$.
    \item Once again, there is log-normal scatter of $\sigma=0.3$ dex around this average CO luminosity for fixed IR luminosity.
\end{itemize}

Using \texttt{limlam\_mocker} once again, we `paint' CO luminosities so modelled onto two separate sets of peak-patch simulations, one for the $z\sim3$ CO(1--0) signal and another for the $z\sim7$ CO(1--0) and CO(2--1) signals.
\begin{itemize}
    \item For the $z\sim3$ signal, we use 9 of the 161 independent peak-patch lightcone simulations generated for Ref.~\cite{Ihle19}. The simulations have a box size of $L_\text{box}=1140$ Mpc and a resolution of $N_\text{cells}=4096^3$. This is sufficient to span 26--34 GHz in CO(1--0) observing frequency and $9.6^\circ\times9.6^\circ$ on the simulated sky, while resolving halos as small as $M_h\simeq2.5\times10^{10}\,M_\odot$ at high completeness. We correct peak-patch halo masses via abundance matching to the halo mass function of~\textcite{Tinker08}. We divide each box into 16 sub-fields of $2^\circ\times2^\circ$ each, for a total of 144 simulations of CO(1--0) at $z\sim3$ as would be observed by COMAP.
    \item For the $z\sim7$ signal, we use 16 lightcones out of a new set of peak-patch simulations designed specifically with reionisation-epoch COMAP signals in mind. These simulations have $L_\text{box}=960$ Mpc and a resolution of $N_\text{cells}=5640^3$, thus spanning $6^\circ\times6^\circ$ on the simulated sky as well as the CO(1--0) and CO(2--1) observing frequencies relevant for COMAP-ERA, while resolving halos slightly below $M_h\simeq10^{10}\,M_\odot$. We again correct peak-patch halo masses via abundance matching to the halo mass function of~\textcite{Tinker08}, incorporating high-redshift corrections from Appendix G of Ref.~\cite{Behroozi13a}. We divide each box into 9 sub-fields of $2^\circ\times2^\circ$ each, for a total of 144 simulations of CO lines at $z\sim7$ as would be observed by COMAP.
\end{itemize}
Here, we apply the two-tier line broadening model of Ref.~\cite{linewidths} to all simulations. All halos with virial mass $M_h>10^{11}\,M_\odot$ are assigned a random inclination $i$ and, based on this $i$ as well as the virial velocity $v_\text{vir}$ of the halo, also assigned a line profile FWHM given by $v_\text{vir}\sin{i}/0.866$. These halos are then binned in 16 linearly spaced bins of line width, while the halos with virial mass $M_h<10^{11}\,M_\odot$ are assumed to have negligible line width compared to the COMAP science channel bandwidth. The CO temperature cube calculated from each bin of halos is convolved with a Gaussian kernel of appropriate scale along the frequency dimension, so that the final line-broadened CO cube is the sum of the Gaussian-filtered 17 CO temperature cubes.

We randomly pair the 144 $z\sim3$ simulations with the 144 $z\sim7$ simulations, ending up with 144 simulated $2^\circ\times2^\circ$ fields. In all cases the Ka-band cube is the sum of the $z\sim3$ CO(1--0) and $z\sim7$ CO(2--1) signals (with the instrument beam and survey transfer function applied to both as detailed for each survey below) with appropriately scaled Gaussian noise, while the Ku-band cube is the sum of the $z\sim7$ CO(1--0) signal (again with the beam and transfer function applied) with appropriately scaled Gaussian noise.

\subsection{COMAP Pathfinder-like survey---reconstructing CO(1--0) with external cross-correlations}
\label{sec:COMAPPF}
As noted above, the principal target for the Pathfinder survey is $z\sim3$ CO(1--0) emission from the epoch of galaxy assembly and peak star formation. Here we consider the possibility of reconstructing this $z\sim3$ signal, with the help of an overlapping galaxy catalogue.

\subsubsection{Methods}

First we further clarify the generation of mock observations. After summing the $z\sim3$ CO(1--0) and $z\sim7$ CO(2--1) signals, we convolve the total Ka-band brightness temperature into a Gaussian filter with the appropriate kernel for the beam size given in~\autoref{tab:comapparams} for the Pathfinder-like observations, as before. However, we will simulate the effect of foreground subtraction in more detail than the simple mean subtraction that we used for simulations in~\autoref{sec:CIILIM}.

Based on the heuristic form of the COMAP transfer function given by Ref.~\cite{COMAPESV}, and a measurement of $\sim70\%$ main beam efficiency used for both COMAP high-redshift and Galactic analyses~\cite{Ihle22,Rennie22}, we approximate the portion of the power spectrum transfer function not included in the Gaussian filter as
\begin{equation}
    \mathcal{T}_\text{hp}(k_\perp,k_\parallel) = \frac{0.49}{(1+e^{5-100\text{\,Mpc}\cdot k_\perp})(1+e^{5-200\text{\,Mpc}\cdot k_\parallel})}.
\end{equation}
We multiply the Fourier transform of the beam-convolved CO cube by the square root of this function, and then invert the Fourier transform to obtain a high-pass filtered signal scaled appropriately by the beam efficiency. These steps transform the signal from a physical brightness temperature into a pseudo-temperature, which for convenience we will notate as $T$. All power spectra in for the remainder of this whole section are implicitly pseudo-power spectra.

By contrast, we will not consider the overlapping galaxy survey in nearly as much detail, given the limited information about each halo in the peak-patch simulations. We mainly select mock `galaxies' from the halos in the $z\sim3$ simulation with virial mass above a certain $M_{h,\text{min}}$, with additional selection possible if desired after that. To gauge the effect of increasing source density and decreasing galaxy bias, we try three different types of selection:
\begin{itemize}
    \item We suppose that the spectroscopic galaxy survey catalogues the locations of all halos with virial mass above $M_{h,\text{min}}/M_\odot=6\times10^{12}$. 
    \item We suppose that the spectroscopic galaxy survey catalogues the locations of all halos with virial mass above $M_{h,\text{min}}/M_\odot=3\times10^{12}$. 
    \item We mock a $z\sim3$ LAE survey similar to the Hobby-Eberly Telescope Dark Energy eXperiment (HETDEX; see Refs.~\citep{HETDEX2008,HETDEX2021inst,HETDEX2021}) by using a slightly modified version of the LAE model of Ref.~\cite{COMAPESV}, which assumed that 5\% of halos with virial mass above $9.3\times10^{10}\,M_\odot$ would be LAE hosts. Here, we first select all halos with virial mass above $M_{h,\text{min}}/M_\odot=10^{11}$, randomly select 5\% of these halos (mimicking a Lyman-alpha emission duty cycle), and then apply a regular mask across the whole survey field that leaves unmasked $50''\times50''$ squares with centres spaced apart by $100''$ in both angular directions. This masking step culls 3/4 of the mock LAEs from the catalogue, and is meant to mimic the sparse sampling of the HETDEX survey, although the actual HETDEX window function will be more complicated. Note that both the 5\% duty cycle and sparse sampling will induce additional shot noise uncorrelated with the CO signal.
\end{itemize}
The resulting source counts per patch for each selection are $718_{-98}^{+103}$, $2872_{-318}^{+312}$, and $5460_{-268}^{+247}$ (90\% intervals) across 144 simulations. Source densities of $\sim10^3$ deg$^{-1}$ per $\Delta z=1$ at this redshift range are eminently plausible from currently operating experiments like the aforementioned HETDEX LAE survey, which expects to leverage integral field spectroscopy catalogue roughly 2000 LAE positions per square degree spanning $z=2$--$3.5$ even with sparse coverage filling only $1/4.5$ of the nominal survey area. Future facilities will also readily achieve such source densities, including ground-based highly multiplexed multifibre spectrographs like DESI-II or MegaMapper targeting $z>2$ Lyman-break galaxies~\citep{MegaMapper,MegaMapperCF}, and the Nancy Grace Roman Space Telescope whose reference survey strategy includes surveying $2\times10^6$ [O\textsc{\,iii}] emitters at $z=2$--3 across 2000 deg$^2$~\citep{RomanHLS}.

We simulate the observations in 48 batches of three fields, mimicking a single COMAP `survey'. Each `survey' shares the same LCB and WF coefficients, designed as follows:
\begin{itemize}
\item We mimic the feed-feed pseudo cross spectrum (FPXS) estimator used by Ref.~\cite{Ihle22} for the CO power spectrum in each field. After generating 38 different mock data cubes with the same signal but with independently generated Gaussian noise with standard deviation per voxel of $\sigma_N\sqrt{38}$ (assumed to be 38 disjoint splits, e.g., in detector and elevation range), we obtain $19\times18$ cross spectra between these cubes and average them to obtain the FPXS estimate for the CO auto spectrum in that field. We then use the average of these FPXS estimators across all fields in the `survey' as our estimate for $P^{(S)}(k)$ purely from the noisy mock data rather than directly from the noiseless input signals. (This method will slightly bias the signal power spectrum due to the common presence of background CO(2--1) emission, but this background is sufficiently subdominant that the extent of over-estimation will be in the range that the LCB filter has already shown to tolerate well in~\autoref{sec:cii_results}.)
\item We obtain the galaxy power spectrum in each field and average across fields. The three-field average estimates the value of $P_1^{(T)}(k)=D_{11}$ required by~\cref{eq:Dijpractical} for the LCB filter.
\item We now average the mock cubes into a single cube with noise per voxel of $\sigma_N$, and obtain cross spectra for each field. The average of the cross spectra across all fields is what we use for $D_{12}=D_{21}=P_{12}^{(S)}(k)$ as required by~\cref{eq:Dijpractical} for the LCB filter.
\end{itemize}
With the relevant auto and cross power spectra estimated for the whole `survey', we apply the LCB and WF reconstructions to each field, measuring reconstruction power spectra as well as normalised cross-correlation against the true (pseudo-)signal on a per-field basis.

\subsubsection{Results}

As before, we note expected detectability briefly. We predict strong detections of CO--galaxy cross spectra with this Pathfinder-like survey. With either smaller catalogues of $\sim700$ mass-selected objects or $\sim5500$ mock LAEs for each 4 deg$^2$ field, the average total (all-$k$) signal-to-noise ratio for the cross power spectrum is 21, rising to 27 for larger mass-selected catalogues of $\sim3000$ objects per field.

The CO(1--0) power spectrum is detectable with a total signal-to-noise ratio of 10 per field on average. This is consistent with the signal-to-noise forecasts from the COMAP collaboration~\citep{COMAPESV} to within a factor of order unity, once we account for various factors such as the number of fields, the inclusion of the contribution of sample variance to noise, differences in voxelisation of the survey volume, approximations made in our transfer function and noise models, calculation details of line broadening, and so on. Again, since the focus of this work is on introducing the reconstruction method rather than forecasting detectability, we consider the level of agreement adequate.

\begin{figure}
    \centering
    \includegraphics[width=0.96\linewidth]{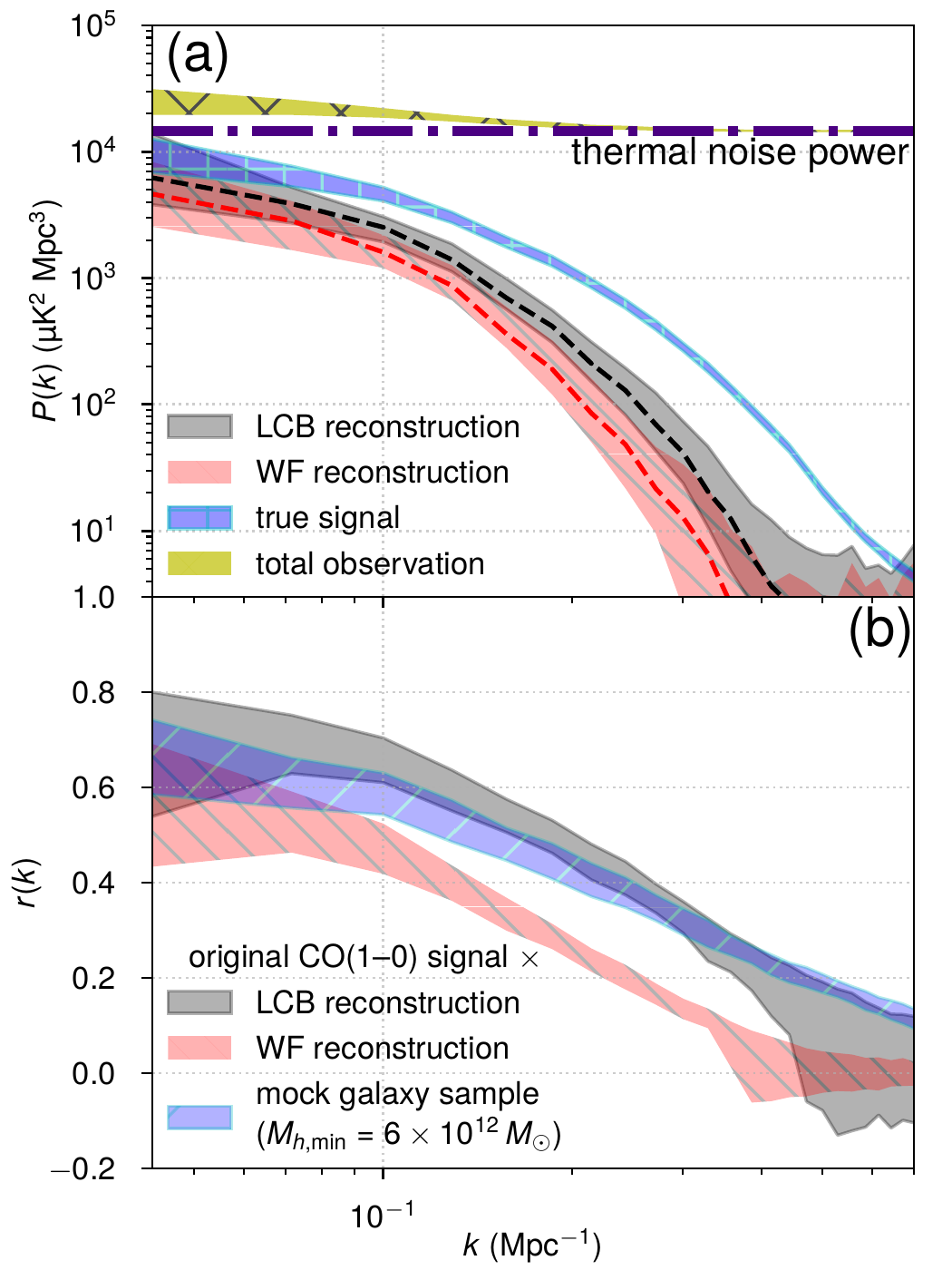}
    \caption{90\% intervals of reconstruction summary statistics across our $144=48\times3$ simulated Pathfinder-like survey fields, given an overlapping galaxy sample simulated with $M_{h,\text{min}}/M_\odot=6\times10^{12}$. We show (a) the obtained power spectra of the LCB and Wiener filter reconstructions alongside their expected values (dashed lines), as well as the original input signal, the total observation including noise and interloper emission, and the thermal noise power spectrum by itself (dash-dotted line). We also show (b) the normalised cross-correlation $r(k)$ between the original input signal and the two reconstructions, as well as between the noiseless CO(1--0) pseudo-signal and the galaxy density contrast.}
    \label{fig:COMAPPF_lessgals}
\end{figure}

\begin{figure}
    \centering
    \includegraphics[width=0.96\linewidth]{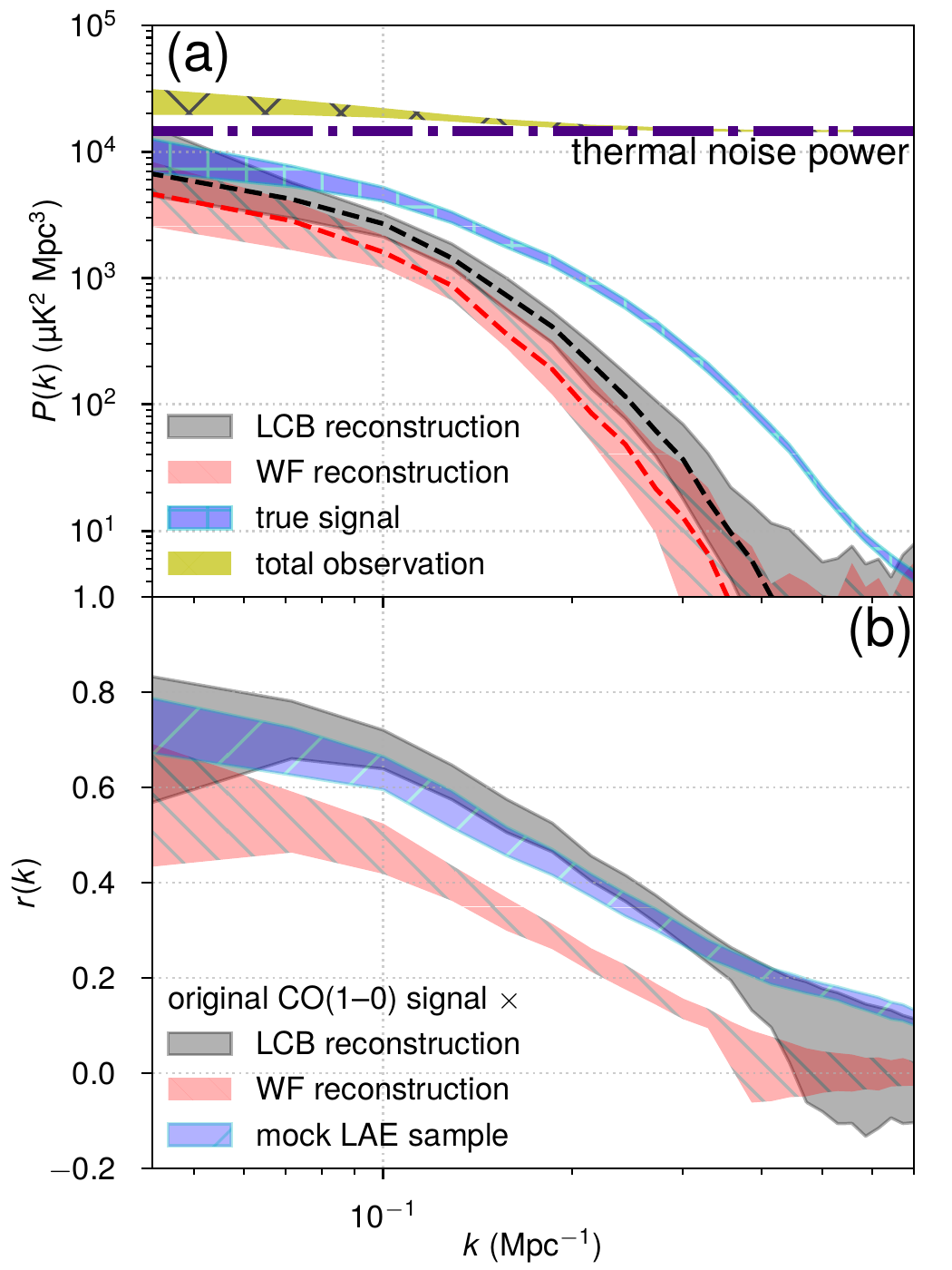}
    \caption{Same as~\cref{fig:COMAPPF_lessgals}, but now for the mock LAE sample with sparse sampling. We again show 90\% intervals across $144=48\times3$ simulated Pathfinder-like survey fields of (a) power spectra for signal, noise, and reconstructions, as well as (b) normalised cross-correlations between the noiseless CO(1--0) pseudo-signal and either a reconstruction or the mock galaxy sample.}
    \label{fig:COMAPPF_LAE_gals}
\end{figure}

\begin{figure}
    \centering
    \includegraphics[width=0.96\linewidth]{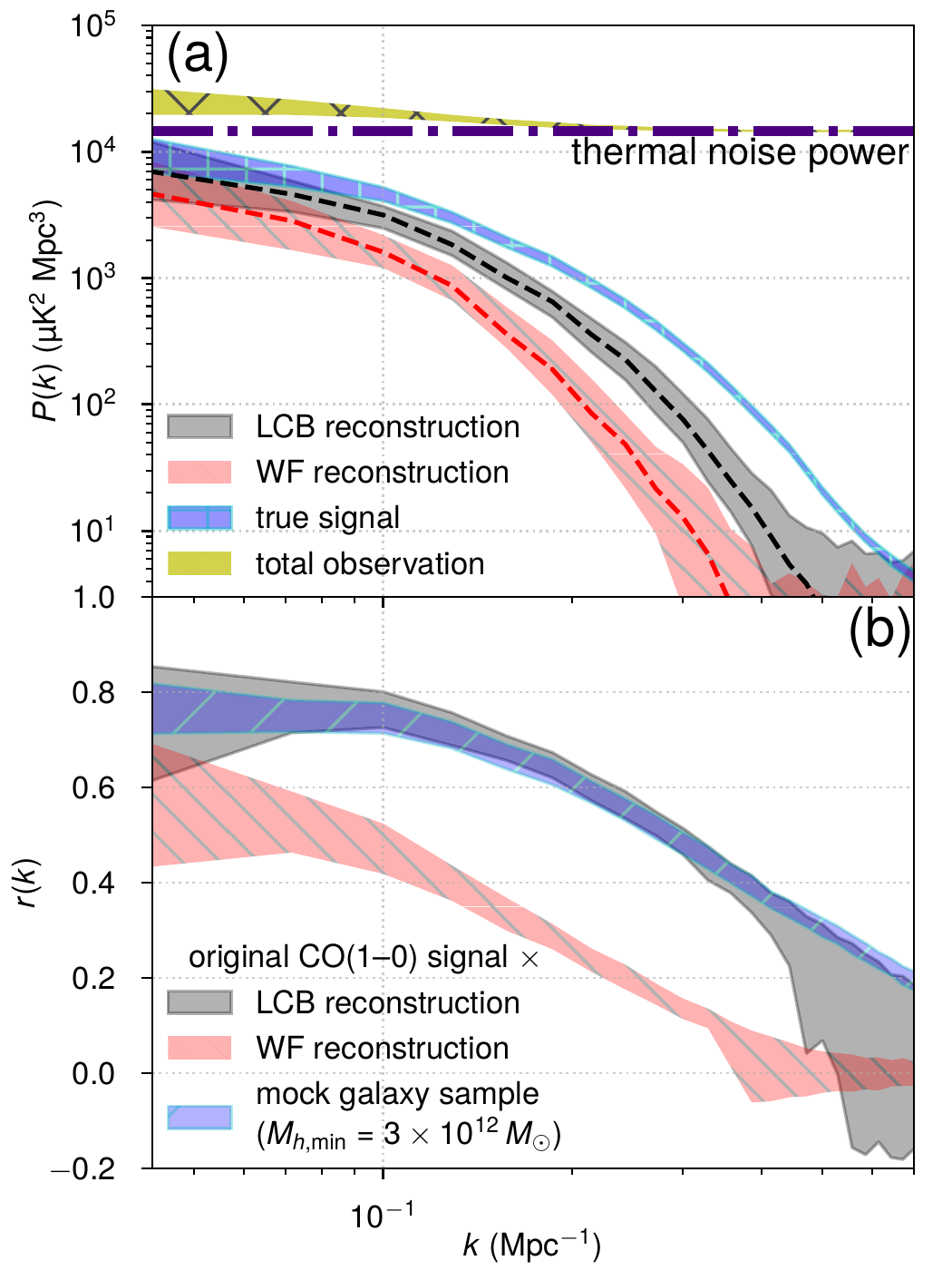}
    \caption{Same as~\cref{fig:COMAPPF_lessgals} or~\cref{fig:COMAPPF_LAE_gals}, but now for an overlapping galaxy sample with $M_{h,\text{min}}/M_\odot=3\times10^{12}$, leading to high source abundances but lower shot noise. We still show 90\% intervals across $144=48\times3$ simulated Pathfinder-like survey fields of (a) power spectra for signal, noise, and reconstructions, as well as (b) normalised cross-correlations between the noiseless CO(1--0) pseudo-signal and either a reconstruction or the mock galaxy sample.}
    \label{fig:COMAPPF_moregals}
\end{figure}

\cref{fig:COMAPPF_lessgals} summarises results for the less abundant mass-selected mock galaxy sample with $M_{h,\text{min}}/M_\odot=6\times10^{12}$. Both LCB and WF reconstructions result in significantly suppressed power spectrum amplitudes, but the LCB filter is able to leverage cross-correlations to significantly improve the normalised cross-correlation against the true pseudo-signal, meaning that we recover the correct structure of fluctuations down to smaller scales. Note that at lower $k$, $r(k)$ is actually slightly higher on average between the LCB reconstruction and the true signal than between the mock galaxies and the true signal. The LCB filter in practice leverages not only the CO--galaxy cross-correlation but also the knowledge of expected structure at large scales, filling in large-scale fluctuations not necessarily completely mapped by our relatively low-density galaxy sample.

Results are similar with our mock LAE sample, as sown in~\cref{fig:COMAPPF_LAE_gals}. Despite a sample just over $7.5$ times more abundant than the $M_{h,\text{min}}/M_\odot=6\times10^{12}$ sample, the lower tracer bias and noise-inducing selection of this sample results in similar detectability and thus similar reconstruction improvement to the lower-abundance sample. But compared to that sample, the mock LAE sample does trace the large-scale modes with higher fidelity, resulting in slightly (although not significantly) improved $r(k)$ and slightly reduced $P(k)$ bias at low $k$. Of course, this in turn slightly increases the advantage over the Wiener filter reconstruction.

The improvement of the LCB filter over the Wiener filter is even more significant with a more abundant mass-selected mock source catalogue with $M_{h,\text{min}}/M_\odot=3\times10^{12}$. \cref{fig:COMAPPF_moregals} shows significantly reduced suppression of the power spectrum of the LCB-filtered data, and even more marked improvement in the normalised cross-correlation, which now stays above 50\% for all $k\lesssim0.3$ Mpc$^{-1}$. The Wiener-filtered data only reaches this $r(k)$ range for $k\lesssim0.1$ Mpc$^{-1}$, where the LCB-filtered data reaches typical $r(k)$ values of 70--80\%. With a more abundant galaxy sample, we also see a reduced difference compared to~\cref{fig:COMAPPF_lessgals} between the normalised cross-correlations of the true signal against the LCB reconstruction versus the galaxy sample. This suggests a scenario more similar to that of~\autoref{sec:CIILIM} where the galaxy sample is sufficiently abundant that the cross-correlation drives the reconstruction at all scales.

\begin{figure*}
    \centering
    \includegraphics[width=0.96\linewidth]{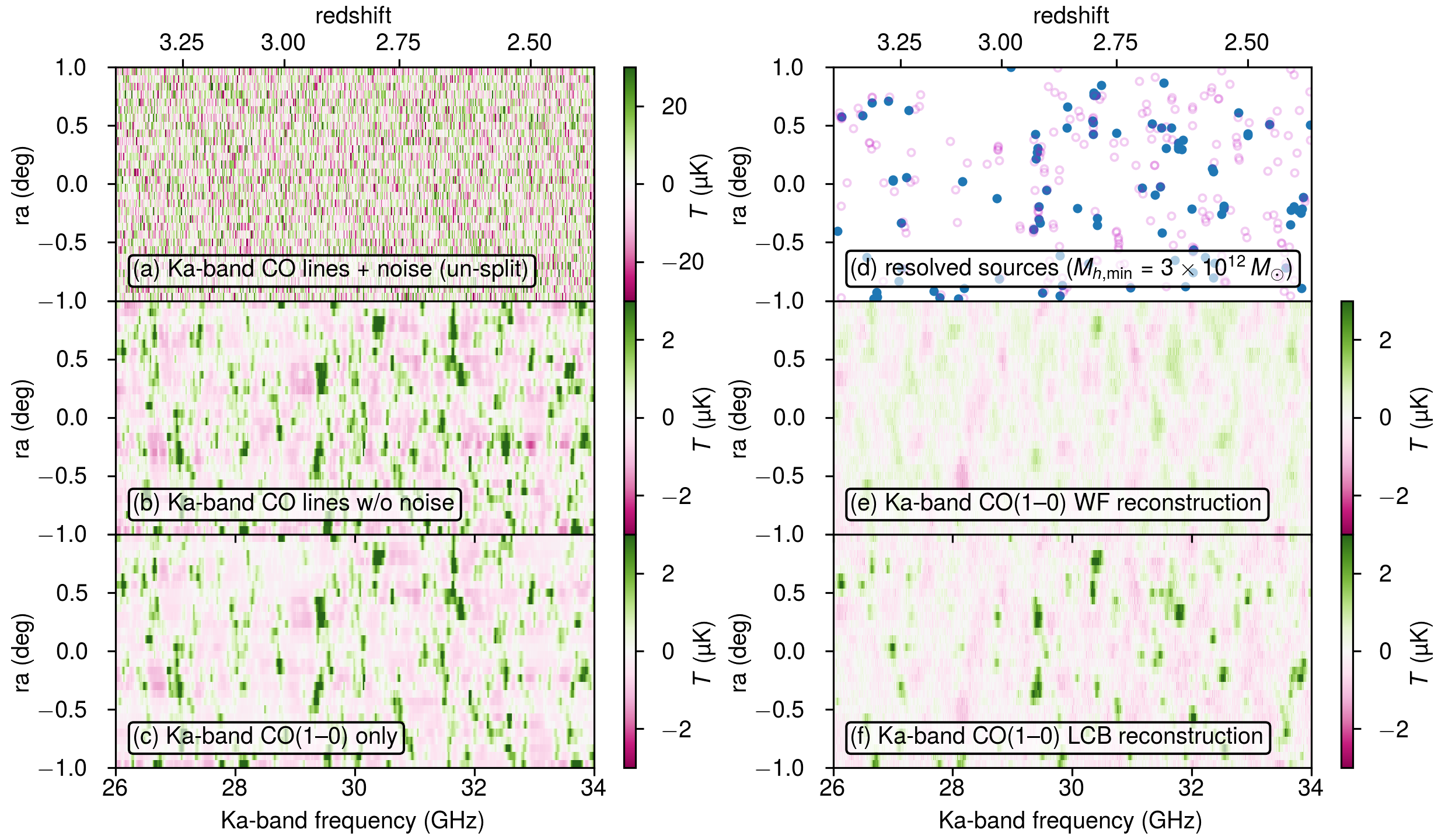}
    \caption{Visualisation of a slice of an example COMAP Pathfinder-like simulation, showing six outputs: (a) the total simulated Ka-band observation, including the CO(1--0) signal from $z\sim3$, the CO(2--1) background from $z\sim7$, and noise; (b) the sum of the noiseless filtered signal and background CO emission; (c) the noiseless $z\sim3$ CO(1--0) pseudo-temperature in isolation; (d) mock galaxy positions contained in the slice shown of the data cube, including both sources properly contained in the slice's declination bin (filled circles) and sources in neighbouring declination bins (unfilled circles); (e) the Wiener-filtered data; and (f) the LCB-filtered data. The mock galaxy sample is generated based on a minimum halo mass of $3\times10^{12}\,M_\odot$.}
    \label{fig:COMAPPF_viz}
\end{figure*}

We conclude consideration of the Pathfinder-like scenario by showing in~\cref{fig:COMAPPF_viz} slices of a typical simulation of signals, observations, and reconstruction, with a source catalogue given by $M_{h,\text{min}}/M_\odot=3\times10^{12}$. Note that the $z\sim3$ CO(1--0) signal clearly dominates over the $z\sim7$ CO(2--1) background, but noise clearly dominates the total observation. As a result, the Wiener filter reconstructs a much weaker set of fluctuations compared to the original signal, whereas the LCB filter as before reconstructs something much closer to the original line-intensity fluctuations where the CO contrast coincides with galaxy density contrast.

\subsection{COMAP-ERA---isolating CO(2--1) with internal cross-correlations}
\label{sec:COMAPER}

The $z\sim3$ CO(1--0) emission dominating over $z\sim7$ CO(2--1) emission presents a boon for the Pathfinder-like survey, but an interloper and a nuisance for COMAP-ERA, which needs to access the reionisation-epoch signal. However, as we discussed in~\autoref{sec:COMAP}, the deployment of lower-frequency instrumentation will allow observations of CO(1--0) at the same redshifts. Thus cross-correlations internal to the COMAP-ERA survey (as opposed to with external datasets) will isolate the common $z\sim7$ CO signal.

\subsubsection{Methods}

We simulate fields in 48 batches of three, thus mocking whole `surveys' at once as in~\autoref{sec:COMAPPF}. In fact, most of the simulation workflow is the same as in~\autoref{sec:COMAPPF}, so we enumerate the main differences.

First, we omit simulation of the FPXS estimation. However, we also `split' the Ku-band data in two, generating two data cubes with the same signal but noise per voxel scaled up by $\sqrt{2}$, and treating them as two independent observations (which they may be at some level if they are obtained from, e.g., different subsets of COMAP-ERA receivers). Thus for the LCB filter we have $n=2$, instead of $n=1$ as in the previous case studies. This Ku-band data `split' would not be useful if we were simply trying to reconstruct the $z\sim7$ CO(1--0) signal as the result would reduce to the Wiener filter as discussed in~\cref{sec:lcbcommon}, but is a handy construction for use with the Ka-band observation to obtain an unbiased estimate of the CO(2--1) power spectrum, as described in~\autoref{sec:lcbbias}. As before, this and all other total auto and cross power spectra are averaged across the fields in each `survey', with the `survey'-wide averages used for LCB filter calculations in each field of the `survey'.

We also adjust the power spectrum transfer function slightly to account for the changed relations between angular or frequency scales with comoving lengths at $z\sim7$:
\begin{equation}
    \mathcal{T}_\text{hp}(k_\perp,k_\parallel) = \frac{0.49}{(1+e^{5-138\text{\,Mpc}\cdot k_\perp})(1+e^{5-144\text{\,Mpc}\cdot k_\parallel})},
\end{equation}
As the transfer function is calculated as a function of $k$, we apply it equally to Ku- and Ka-band CO cubes.

For $z\sim7$ CO(2--1), we offer the Wiener filter reconstruction the best possible chance to succeed by using the \emph{true signal pseudo-power spectrum for each field} with the total Ka-band auto power spectrum to Wiener-filter the data. However, this advantage is entirely irrelevant in the presence of strong interloper emission, which the results of~\autoref{sec:cii_results} should already suggest.

We also cut Fourier modes with $k<0.01$ or $k>0.7$ from the LCB reconstruction, as we expect cross-correlation information at those scales to be spurious at best due to the COMAP beam and pipeline transfer functions simulated here. This step is not necessary for the Wiener filter, as again we provide it with the true signal power spectrum.

\subsubsection{Results}

We briefly advertise average signal-to-noise ratio expectations, which are 52 for the Ku-band CO(1--0) signal, 12 for the Ka-band CO(2--1) signal, and 28 for the cross-band CO power spectrum. All these values are broadly in line with the projections of~\cite{Breysse22} for the same CO emission model, and suggest that the cross power spectrum has significant information to aid in reconstruction with the LCB filter.

\begin{figure}
    \centering
    \includegraphics[width=0.96\linewidth]{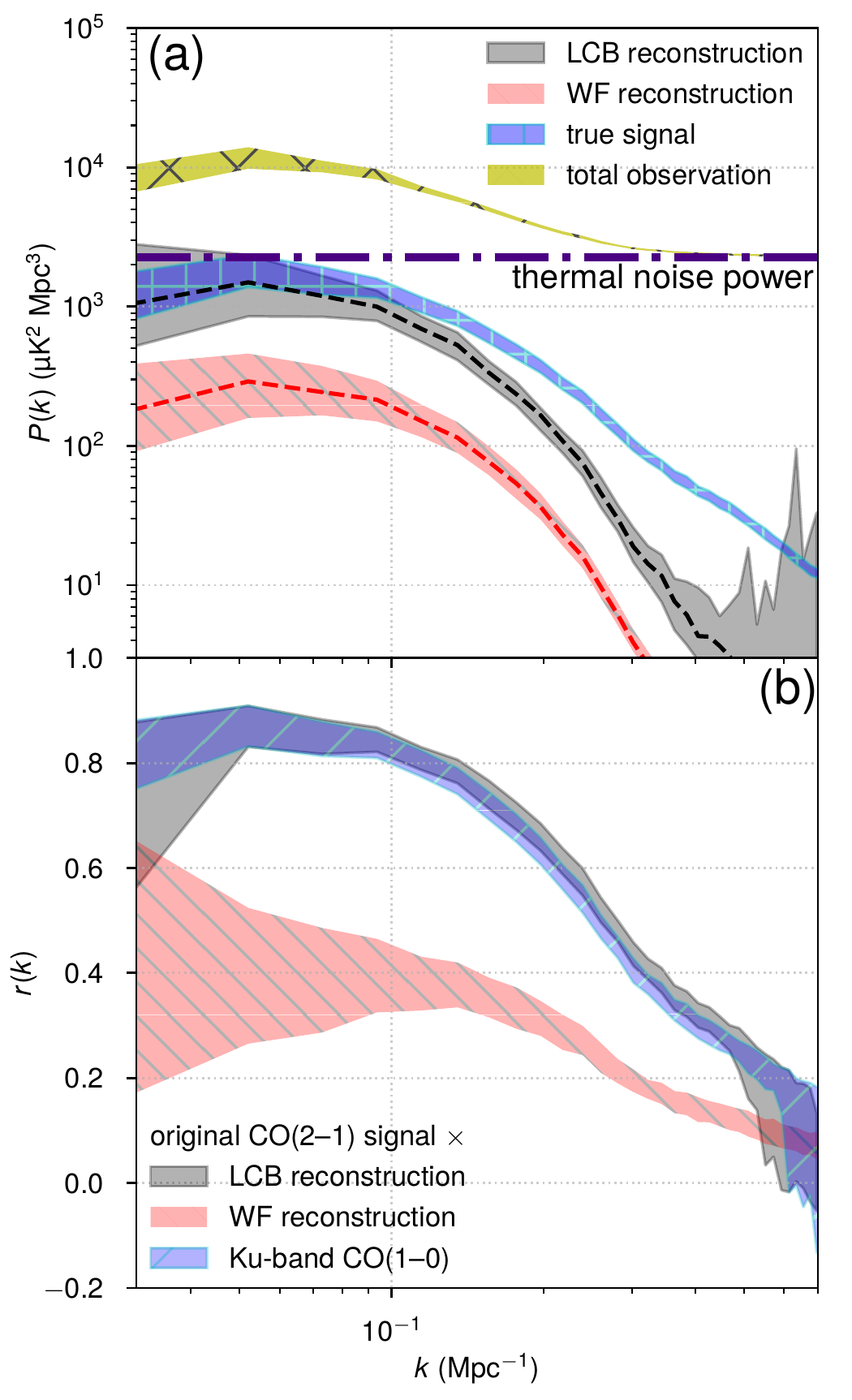}
    \caption{90\% intervals of reconstruction summary statistics for Ka-band data across our $144=48\times3$ simulated COMAP-ERA survey fields. We show (a) the obtained power spectra of the LCB and Wiener filter reconstructions alongside their expected values (dashed lines), as well as the original input signal, the total Ka-band observation including noise and interloper emission, and the thermal noise power spectrum by itself (dash-dotted line). We also show (b) the normalised cross-correlation $r(k)$ between the original input signal and the two reconstructions, as well as between the noiseless CO(1--0) pseudo-signal and the galaxy density contrast.}
    \label{fig:COMAPERA}
\end{figure}

\cref{fig:COMAPERA} shows the key results across our simulations. The qualitative results are largely the same as for the [C\textsc{\,ii}] survey simulated in~\autoref{sec:cii_results}. The principal difference is that here the cross-correlation target is not an external sample of discrete galaxies, but other line-intensity mapping data. This distinction is irrelevant to the LCB filter, which recovers the lowest-$k$ modes with $r(k)$ exceeding 80\% and generally recovers linear-scale CO(2--1) fluctuations much better than the Wiener filter.

\subsubsection{Limitations of reconstructions}

At this point, we have repeatedly established the ability of the LCB filter to robustly reconstruct the true signal with significant improvements over the Wiener filter. We now demonstrate some limitations of these reconstructions, in the context of examining the relation between CO(2--1) and CO(1--0) emission surveyed by COMAP-ERA across the same volume.

The CO(2--1)/CO(1--0) intensity ratio is a key probe of the environments traced by CO-bright molecular gas, being associated with the temperatures and densities of molecular clouds as well as heating from the interstellar radiation field, and the consequent excitation of the CO rotational transitions~\citep{Hasegawa97,Hasegawa97b,Sorai01,Koda12,Koda20,denBrok21}. A global measurement of this line ratio should be a statistical diagnostic for the dynamics and chemistry of the earliest molecular clouds and galactic dust in cosmic history, and would perhaps improve prospects of recovering quantities like the cosmic molecular gas density in a physically motivated fashion.

Since LIM analyses will typically discard information about the mean intensity in cleaning the data of continuum foregrounds, the best that one can infer is the ratio of CO(2--1) temperature fluctuations to CO(1--0) temperature fluctuations. For fluctuations on linear scales, the result will not be the ratio of the line temperatures, but the ratio of the line temperature--bias products $\avg{Tb}_\text{line}$, which we briefly mentioned in~\autoref{sec:lcbbias} as the conversion between matter density contrast and CO temperature contrast. That said, in principle we expect these lines to correlate strongly (see, e.g.,~\cite{Yang21b}) and trace the underlying matter density with similar biases. This expectation is certainly the case for the CO model used in this work. At $z=6.68$ (corresponding to the midpoints of the COMAP observing frequency bands), the CO(2--1)/CO(1--0) ratio in $\avg{Tb}$ (calculated via~\cref{eq:Tbdef}, assuming the halo bias model of Ref.~\cite{Tinker10} and the same halo mass function used for peak-patch mass adjustments in~\autoref{sec:CO_sims}) is 0.74. The CO(2--1)/CO(1--0) ratio in $\avg{T}$, calculated simply by removing $b(M_h)$ from the integrand of~\cref{eq:Tbdef}, is only around 5\% lower at 0.71.

\begin{figure}
    \centering
    \includegraphics[width=0.96\linewidth]{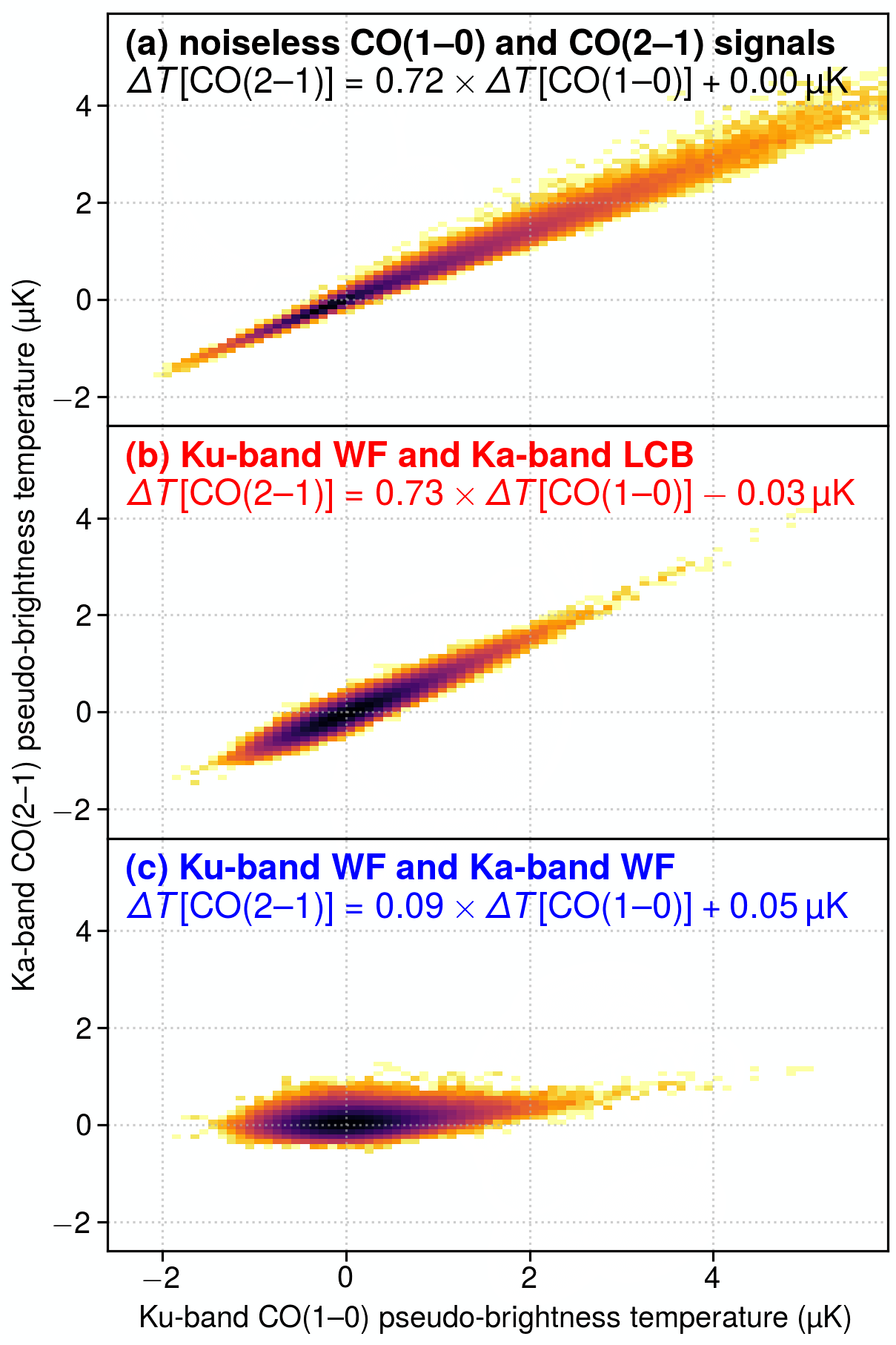}
    \caption{Joint probability distributions, shown using a log colour scale (darker colours indicate higher probability), of CO(1--0) and CO(2--1) temperature fluctuations at $z\sim7$. We show distributions for a representative realisation of (a) the pair of noiseless pseudo-signals, (b) simulated COMAP-ERA reconstructions using the Wiener filter or the LCB filter as appropriate, and (c) simulated COMAP-ERA reconstructions using only Wiener filters for each band. The LCB reconstruction, although failing to help recover the correct shape of the joint probability distribution, allows recovery of the correct slope of the correlation and thus the correct global line ratio in $\avg{Tb}$ to within a few percent, modulo bias from observational distortions. Using only Wiener filter reconstructions fails to recover even the correct slope of the correlation line.}
    \label{fig:COMAPERAratio}
\end{figure}

We should be able to find the CO(2--1)/CO(1--0) ratio in $\avg{Tb}$ from simulated COMAP-ERA data in two ways. The first is by taking the ratio of the CO(1--0)--CO(2--1) cross power spectrum to the CO(1--0) auto power spectrum on linear scales. Simulated surveys recover a ratio of $0.75\pm0.03$ (68\% interval) on average across $48\times3$ patches, so recovery is successful and unbiased. The second is by finding a linear fit to the reconstructed pseudo-temperature fields, namely the LCB-filtered Ka-band CO(2--1) data and the Wiener-filtered Ku-band CO(1--0) data. We show the correlation between the reconstructions in~\cref{fig:COMAPERAratio} through their joint probability distribution, alongside the actual correlation of the original, noiseless pseudo-temperature fields.

The original pseudo-signals have a CO(2--1)/CO(1--0) ratio of 0.72 in $\avg{Tb}$ with very little scatter from realisation to realisation, showing a slight bias relative to the true value resulting from observational effects (in particular the smaller beam FWHM and thus reduced beam dilution in the Ku-band observation compared to the Ka-band observation). But once we accept this bias between the true ratio and the pseudo-signal ratio, we note that the reconstructions (which, again, are effectively \emph{pseudo}-pseudo-signals) successfully recover the correct pseudo-signal ratio, obtaining $0.73\pm0.02$ (68\% interval) across our $48\times3$ simulated patches. Using only Wiener filter reconstructions naturally fails to recover the same line ratio, due to strong suppression of the Ka-band reconstruction by the overwhelming $z\sim3$ interloper emission.

But while we recover the correct overall slope of the correlation, the reconstructions evidently do not recover the actual shape of the joint probability distribution. The original pseudo-signals show a tight correlation at low pseudo-temperatures that broadens for the brightest CO peaks. These trends are sensible given our model, which assumes a log-normal scatter in CO(1--0) or CO(2--1) luminosity for fixed star-formation rate. Because the scatter is log-normal, at low luminosities and thus for small temperature fluctuations, the absolute difference between the CO(1--0) and CO(2--1) signals will be smaller than for high luminosities and large temperature fluctuations. The original joint probability distribution, even with distortion by observational effects, thus contains interesting information about the stochasticity of CO emission and its relation to CO luminosity. No such information is recovered by the reconstructions, which do show skewed distributions but not the correct size or shape of skew. This shortcoming is to be expected from applying a linear estimator to two highly non-Gaussian fields. In this context, the LCB and Wiener filters recover large-scale fluctuations well, but small-scale statistics less so.

Ultimately, the LCB filter is valuable in translating the cross-correlation between correlated signals into a redshift-space map that robustly rejects disjoint systematics and noise, but will not recover information about global properties beyond the power spectra used to calculate the filter. More sophisticated reconstruction techniques, potentially with sensible informative inbuilt priors, would likely significantly improve recovery of local statistics of and between observables.
\begin{figure*}
    \centering
    \includegraphics[width=0.96\linewidth]{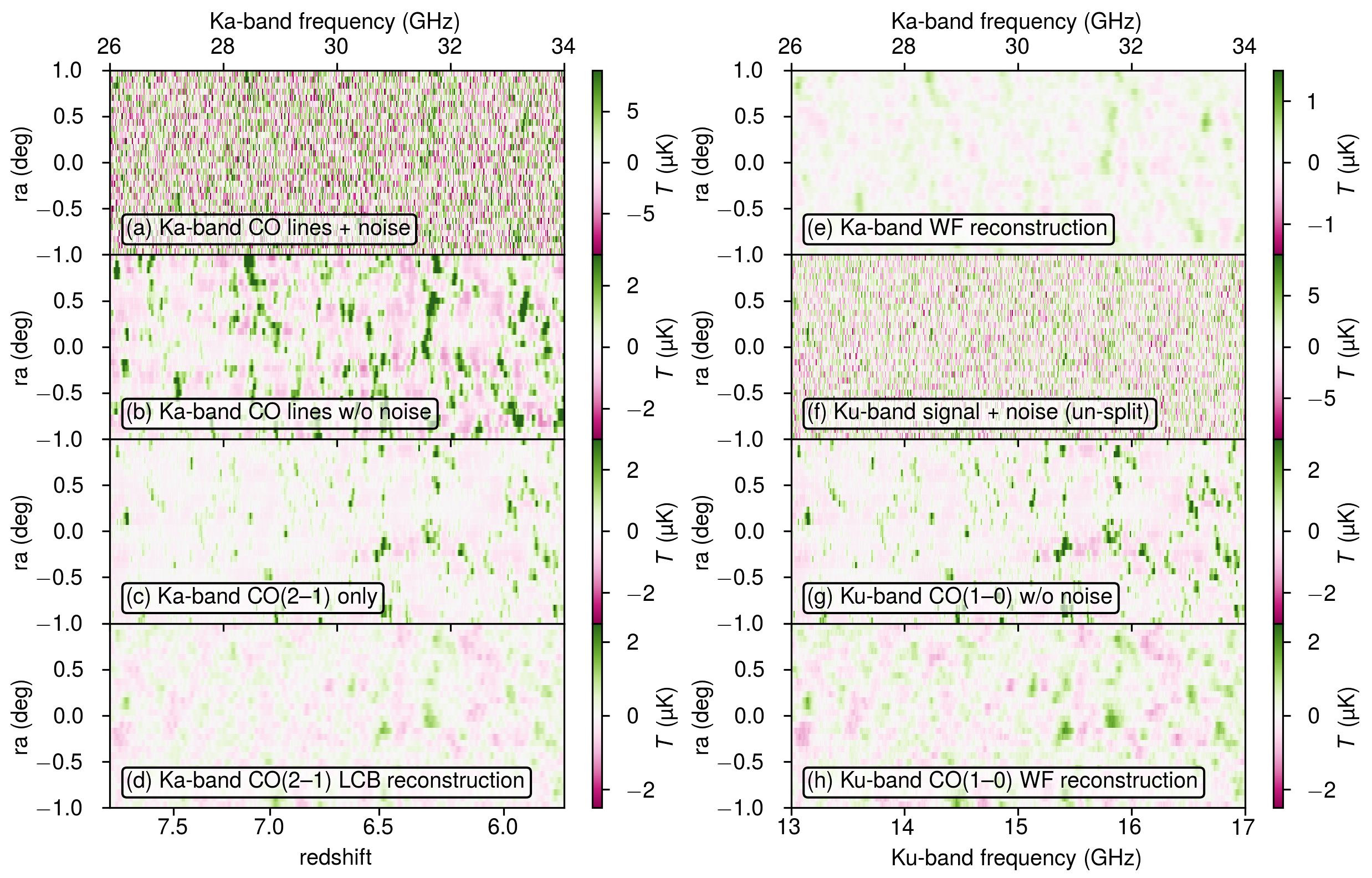}
    \caption{Visualisation of a slice of an example COMAP-ERA simulation, showing six outputs: (a) the total simulated Ka-band observation, including the CO(1--0) interloper from $z\sim3$, the CO(2--1) signal from $z\sim7$, and noise; (b) the sum of the noiseless filtered signal and interloper CO emission; (c) the noiseless $z\sim7$ CO(2--1) pseudo-temperature in isolation; (d) the LCB-filtered Ka-band data, based on cross-correlation with the Ku-band measurement of CO(1--0); (e) the Wiener-filtered Ka-band data; (f) the total simulated Ku-band observation including noise; (g) the noiseless Ku-band CO(1--0) signal; and (h) the Wiener-filtered Ku-band data.}
    \label{fig:COMAPERAviz}
\end{figure*}

In concluding discussion of COMAP-ERA simulation results, we plot slices of various mock observations and reconstructions from a typical realisation of a survey field in~\autoref{fig:COMAPERAviz}, serving as a qualitative graphical recap.
\section{Conclusions}
\label{sec:conclusions}
We can now decisively answer the questions we posed in the~\hyperref[sec:intro]{Introduction}: \emph{Can the LCB filter successfully reconstruct the structure of line-intensity fluctuations on large linear scales by exploiting}
\begin{itemize}
    \item \emph{cross-correlations with surveys of discrete sources?} Yes. We showed this both in the context of a futuristic [C\textsc{\,ii}] LIM survey and in the context of near-term observations feasible with the COMAP Pathfinder. Both normalised cross-correlation and estimator power spectrum bias relative to the true signal improve significantly over the Wiener filter, across a range of different assumptions about the information available to calculate the filter.
    \item \emph{cross-correlation with a measurement of a different, correlated line-intensity signal?} Yes. Simulations of the proposed COMAP-ERA survey show strong recovery of $z\sim7$ CO(2--1) from Ka-band data through cross-correlation with Ku-band data measuring $z\sim7$ CO(1--0), despite the presence of bright interloper emission in CO(1--0) from $z\sim3$ in the Ka band. The LCB filter should thus enable COMAP-ERA to robustly map molecular gas in multiple lines across large volumes spanning the late epoch of reionisation.
\end{itemize}

As we have stated multiple times---as early as the~\hyperref[sec:intro]{Introduction}, in fact---the LCB filter is not necessarily the best estimator to use, given the strong non-Gaussianity of a line-intensity field that principally traces the gas content of galaxies. We saw this in the COMAP-ERA linear reconstructions, which failed to recover the correct skew of the joint probability distribution between CO(1--0) and CO(2--1) temperature fluctuations. Furthermore, the shot noise components of different tracers could become a significant source of small-scale reconstruction variance in this formalism. However, at large linear scales where the signal \emph{is} approximately Gaussian, the LCB filter appears successful, suggesting that it or a similarly straightforward technique may be sufficient for some use cases. For instance, we hope future work will investigate identification of large ($\gtrsim10$ Mpc \replaced{radius}{diameter}) voids and peaks in the context of COMAP-ERA reconstructions. \added{(A detailed consideration is beyond the scope of this work but a preliminary consideration of the [C\textsc{\,ii}] simulations from~\autoref{sec:CIILIM}, as outlined in~\autoref{sec:peakfinding} suggests reason for optimism.) }Factoring in redshift-space distortions and anisotropic information, which we have neglected here, would also likely improve reconstruction.

We also hope that the possibility of robust signal reconstruction through cross-correlation further motivates coordination between future LIM surveys \emph{and} between LIM and other LSS surveys. Such collaboration will enable looking beyond statistical constraints on the early cosmic history of star formation and galaxy assembly, towards directly mapping these processes across large cosmic volumes in many different shades.
\begin{acknowledgments}
Thanks to Dick Bond, Patrick Breysse, Ren\'{e}e Hlo\v{z}ek, Zack Li, Hamsa Padmanabhan, and others at the University of Toronto and/or in the COMAP collaboration for discussions that encouraged the writing of this paper. Thanks also to George Stein for running and making available the original peak-patch simulations for Ref.~\cite{Ihle19}.\added{ DTC also thanks an anonymous referee who provided thoughtful comments and questions during the review of this paper.}

Research in Canada is supported by NSERC and CIFAR. Parts of this work were performed on the GPC and Niagara supercomputers at the SciNet HPC Consortium. SciNet is funded by: the Canada Foundation for Innovation under the auspices of Compute Canada; the Government of Ontario; Ontario Research Fund---Research Excellence; and the University of Toronto.

Parts of this work make use of UniverseMachine-processed snapshots of the SMDPL simulation. The author is grateful to Peter Behroozi for making the UniverseMachine code and catalogues publicly available. As for the original SMDPL simulation, the author acknowledges the Gauss Centre for Supercomputing e.V.~and the Partnership for Advanced Supercomputing in Europe (PRACE) for funding the MultiDark simulation project by providing computing time on the GCS Supercomputer SuperMUC at Leibniz Supercomputing Centre (LRZ).

DTC is supported by a CITA/Dunlap Institute postdoctoral fellowship. The Dunlap Institute is funded through an endowment established by the David Dunlap family and the University of Toronto. The University of Toronto operates on the traditional land of the Huron-Wendat, the Seneca, and the Mississaugas of the Credit River; DTC is grateful to have the opportunity to work on this land.\added{ DTC also acknowledges support through the Vincent and Beatrice Tremaine Postdoctoral Fellowship at CITA.}
\end{acknowledgments}

\appendix
\section{Proof of equivalence of LCB filtering with the estimator of \texorpdfstring{\textcite{ManzottiDodelson14}}{Manzotti and Dodelson [2]}}
\label{sec:lcb_md14}
We show here explicitly that the LCB filter is equivalent to the optimal map estimator of~\textcite{ManzottiDodelson14}. Although this was shown explicitly for $n=1$ by~\textcite{Weaverdyck18}, we show it for all $n$, with the only assumption being that there is zero covariance between different wavenumbers $k$, or between spherical harmonic multipoles $\ell$ in the use case of~\textcite{ManzottiDodelson14}.

First, we note that the matrix $D$ is defined in much the same way as we define it in this work, e.g., in~\autoref{eq:Dijdef}. However, they allocate the observable of interest to index 1; we will adapt any of their expressions to our convention of allocating index $p=n+1$. There is an additional covariance for which~\textcite{ManzottiDodelson14} use the notation $C$, but to denote the covariance of $d_p-s_p$, i.e., the sum of the covariances of all components of the observable $p$ that we are \emph{not} interested in reconstructing (noise, other uncorrelated interlopers, and so on). So their $C$ is really our $P^{(N)}$.

\textcite{ManzottiDodelson14} also define a further additional variance $N$, and we reproduce the definition of their Eq.~(8) here (albeit with necessary adaptations as described above):
\begin{align}
    N^{-1}=\left(P^{(N)}\right)^{-1} + \left(D^{-1}\right)_{pp}.\label{eq:MD14N}
\end{align}
Given this $N$, we now reproduce the estimator of~\textcite{ManzottiDodelson14} as defined by their Eq.~(9) (with some adaptations, and omitting the dependence on $k$ for brevity):
\begin{align}
    \hat{s}_p=N\left[\left(P^{(N)}\right)^{-1}d_p-\sum_{j=1}^n(D^{-1})_{pj}d_j\right].\label{eq:MD14est}
\end{align}
First, recall that our formalism involves a Cholesky decomposition of $D$ such that $D=LL^T$, with $L$ being a lower triangular matrix such that $L_{ij}=0$ if $i<j$. Then since $L^{-1}$ is also a lower triangular matrix (and its transpose $L^{-T}$ is upper triangular),
\begin{align}
    1=(L^{-1}L)_{pp}=\sum_{i=1}^p(L^{-1})_{pi}L_{ip}=(L^{-1})_{pp}L_{pp},
\end{align}
and
\begin{align}
    (D^{-1})_{pp}&=\sum_{i=1}^p(L^{-T})_{pi}(L^{-1})_{ip}\nonumber\\&=\sum_{i=1}^p(L^{-1})_{ip}^2=(L^{-1})_{pp}^2=(L_{pp})^{-2}.\label{eq:Dinvpp}
\end{align}
More generally,
\begin{align}
    (D^{-1})_{pj} &= \sum_{i=1}^p(L^{-T})_{pi}(L^{-1})_{ij}=(L^{-1})_{pp}(L^{-1})_{pj}\nonumber\\&=(L_{pp})^{-1}(L^{-1})_{pj}.\label{eq:Dinvpj}
\end{align}
Additionally, since $LL^{-1}=I$ (the identity matrix),
\begin{align}
    \delta_{pj}&=(LL^{-1})_{pj}=\sum_{i=1}^pL_{pi}(L^{-1})_{ij}\nonumber\\&=L_{pp}(L^{-1})_{pj}+\sum_{i=1}^nL_{pi}(L^{-1})_{ij}.\label{eq:trickery}
\end{align}
Then for all $j<p$, i.e., for all $j$ from 1 to $n$, putting together~\cref{eq:Dinvpj} and~\cref{eq:trickery} gives us
\begin{align}
    (D^{-1})_{pj}=-\frac{1}{L_{pp}^{2}}\sum_{i=1}^nL_{pi}(L^{-1})_{ij}.\label{eq:trickery2}
\end{align}
Substituting this result,~\cref{eq:MD14N}, and~\cref{eq:Dinvpp} all at once into~\cref{eq:MD14est}, we find
\begin{align}
    \hat{s}_p&=\frac{L_{pp}^2P^{(N)}}{L_{pp}^2+P^{(N)}}\left(\frac{d_p}{P^{(N)}}+\frac{1}{L_{pp}^2}\sum_{j=1}^n\sum_{i=1}^nL_{pi}(L^{-1})_{ij}d_j\right)\nonumber\\&=\frac{L_{pp}^2}{L_{pp}^2+P^{(N)}}d_p+\frac{P^{(N)}}{L_{pp}^2+P^{(N)}}\sum_{j=1}^n\sum_{i=1}^nL_{pi}(L^{-1})_{ij}d_j.
\end{align}
One may readily see that this is entirely equivalent to~\autoref{eq:LCB}, which is what was to be shown.

\section{Further expressions for use with the LCB filter when \texorpdfstring{$n>1$}{n>1}}
\label{sec:lcbmaths}
It is straightforward to show that in general, each element of the $i$th row of the Cholesky decomposition of $D$ may be expressed recursively in terms of the $(i-1)\times(i-1)$ submatrix preceding it as well as the preceding elements of that row:
\begin{align}
    L_{ii}&=\sqrt{D_{ii}-\sum_{j=1}^{i-1}L_{ij}^2};\\
    L_{ij}&=\frac{1}{L_{jj}}\left(D_{ij}-\sum_{k=1}^{j-1}L_{ik}L_{jk}\right),
\end{align}
where the latter is for $j<i$ only.

Specifically in the case of $n=2$, where $D$ is a $3\times3$ matrix, we obtain the following elements of $L$:
\begin{align}
    L_{11}&=\sqrt{D_{11}};\\
    L_{21}&=D_{21}/\sqrt{D_{11}};\\
    L_{22}&=\sqrt{D_{22}-D_{21}^2/D_{11}};\\
    L_{31}&=D_{31}/\sqrt{D_{11}};\\
    L_{32}&=(D_{32}-D_{21}D_{31}/D_{11})/L_{22};\\
    L_{33}&=\sqrt{D_{33}-\frac{D_{31}^2}{D_{11}}-L_{32}^2}.
\end{align}
Note that we do not expand $L_{22}$ and $L_{32}$ in the expressions for $L_{32}$ and $L_{33}$ respectively, for brevity. It should be clear that the first three of these expressions are the same as for a $2\times2$ matrix, and evaluate to~\cref{eq:chol11n1,eq:chol12n1,eq:chol22n1} given $D$ as defined in~\cref{eq:Dijdef}.

In particular, note that since
\begin{align}
    L_{pp}^2&= D_{pp}-\sum_{j=1}^{n}L_{pj}^2 = P^{(S)}-\sum_{j=1}^{n}L_{pj}^2,
\end{align}
it is always the case that $L_{pp}^2+P^{(N)}$ can be written purely in terms of the total power spectrum $P^{(S)}+P^{(N)}$ of $d_p$ and the upper left $n\times n$ submatrix of $L$, which in turn only depends on the total power spectra of the other observables.

\section{Equivalence of the LCB filter for \texorpdfstring{$s_i=s_j$}{si=sj} for all \texorpdfstring{$i$}{i} and \texorpdfstring{$j$}{j} with a Wiener filter for the inverse noise variance-weighted average of all data}
\label{sec:lcbcommon}

Suppose now that all our observations are of the same underlying signal, i.e., $s_i=s_j$ for all $i$ and $j$, and call this signal $s\equiv s_i$ for all $i$. The observations still have independent, uncorrelated noise, possibly even different levels of noise. However, importantly, the expected values of the cross power spectra are the same as the expected value of the signal auto power spectrum, unbiased by noise. So the relevant covariance matrix is given by
\begin{equation}
    D_{ij}(k) = \begin{cases}P^{(S)}(k) + \delta_{ij}P_i^{(N)}&\text{if }i\leq n\text{ and }j\leq n,\\P^{(S)}(k)&\text{otherwise.}\end{cases}\label{eq:Dijdefn1}
\end{equation}
For intuition, consider again the special case of $n=1$. Then~\cref{eq:LCBn1ratio} becomes
\begin{equation}
\frac{L_{22}^2}{L_{22}^2+P_2^{(N)}}=\frac{P^{(S)}P_1^{(N)}}{\left(P^{(S)}+P_1^{(N)}\right)\left(P^{(S)}+P_2^{(N)}\right)-{P^{(S)}}^2}.\label{eq:LCBn1sratio}
\end{equation}
and we can rewrite~\cref{eq:LCBn1} as
\begin{align}
    \hat{s} &= \frac{P^{(S)}\left({P_1^{(N)}}d_2+{P_2^{(N)}}d_1\right)}{P^{(S)}\left(P_1^{(N)}+P_2^{(N)}\right)+P_1^{(N)}P_2^{(N)}}\nonumber\\&=
    \frac{P^{(S)}d_*}{P^{(S)}+P^{(N*)}},\label{eq:LCBn1s}
\end{align}
where $d_*$ denotes the noise inverse variance-weighted average
\begin{equation}
    d_*=\frac{d_1/P_1^{(N)}+d_2/P_2^{(N)}}{1/P_1^{(N)}+1/P_2^{(N)}},
\end{equation}
whose power spectrum will be the sum of the signal power spectrum, which will still be $P^{(S)}$, and a noise power spectrum that one may readily find equal to
\begin{align}
    P^{(N*)}\equiv\frac{1}{1/P_1^{(N)}+1/P_2^{(N)}}.
\end{align}
Written in this way, we can easily see that the LCB-filtered estimator for the signal from the covariance defined in~\cref{eq:Dijdefn1} is equivalent to applying a Wiener filter to the inverse variance-weighted $d_*$ in the case of $n=1$.

Now consider the more general case. We will consider the estimator in the form given by~\textcite{ManzottiDodelson14}, which we handily proved equivalent to the LCB filter in~\cref{sec:lcb_md14}. It is straightforward to find the inverse of the covariance matrix $D$, given its simple form. Namely, the upper left $n\times n$ submatrix of $D^{-1}$ is given by
\begin{equation}
    (D^{-1})_{ij} = \delta_{ij}/P_i^{(N)},
\end{equation}
for $i\leq n\text{ and }j\leq n$. Meanwhile, for $i<p$, the final row and final column are both given by
\begin{equation}
    (D^{-1})_{jp} = (D^{-1})_{pj} = -1/P_j^{(N)},
\end{equation}
and finally,
\begin{equation}
    (D^{-1})_{pp} = 1/P^{(S)}+\sum_{j=1}^n 1/P_{j}^{(N)}.
\end{equation}
Note that the upper left $n\times n$ submatrix is completely irrelevant to the estimator, but we have provided it here for completeness.

In any case, the variance $N$ associated with $\hat{s}_p$ as given by~\cref{eq:MD14N} is
\begin{equation}
    N^{-1} = 1/P^{(N)}+(D^{-1})_{pp} = 1/P^{(S)}+\sum_{j=1}^p 1/P_{j}^{(N)},
\end{equation}
such that the estimator as given by~\cref{eq:MD14est} evaluates to
\begin{align}
    \hat{s}_p=\frac{\sum_{j=1}^pd_j/P_j^{(N)}}{1/P^{(S)}+\sum_{i=j}^p 1/P_{j}^{(N)}}.
\end{align}
But in this general case,
\begin{equation}
    d_* = \frac{\sum_{j=1}^pd_j/P_j^{(N)}}{\sum_{i=j}^p 1/P_{j}^{(N)}},
\end{equation}
and
\begin{equation}
    1/P^{(N*)} = \sum_{i=j}^p 1/P_{j}^{(N)},
\end{equation}
so that the estimator is
\begin{align}
    \hat{s}_p=\frac{d_*/P^{(N*)}}{1/P^{(S)}+1/P^{(N*)}}=\frac{P^{(S)}d_*}{P^{(S)}+P^{(N*)}},
\end{align}
which is equivalent to applying a Wiener filter to the inverse variance-weighted $d_*$, just as we sought to show for all $n$. This is hardly surprising as both the Wiener filter and the LCB filter are optimal map estimators, and it should not be possible to obtain extra information from the same data simply by cross-correlating splits.

\section{A preliminary consideration of improvements in peak identification with LCB reconstructions}
\label{sec:peakfinding}

\added{Due to the varied ways in which one may identify and qualify peaks and voids in the cosmic web, a detailed consideration of improvements in peak or void identification is well beyond the scope of the present work. However, with an extremely basic peak identification workflow, we may demonstrate probable cause to expect a significant improvement in the ability of a survey to identify line-intensity peaks through a LCB reconstruction when compared to the Wiener-filtered reconstruction.

For this, we return to the 10 SMDPL/UniverseMachine lightcones and their corresponding [C\textsc{\,ii}] simulations used for~\autoref{sec:CIILIM}. Immediately, based on the size of the simulation, we note that the survey only measures a limited number of modes at scales of $\sim100$ Mpc, so we discard modes with $k<0.05$ Mpc$^{-1}$ from both the LCB and Wiener filter reconstructions. We also blur the data cube with a 3D Gaussian filter with its profile defined by a comoving RMS width of $4.2$ Mpc in all directions. This corresponds to a FWHM of 10 Mpc, reflecting our qualitative conclusion that the fidelity of the LCB reconstruction is best suited to identification of $\gtrsim10$ Mpc diameter features.

From this preprocessed data cube we identify the brightest voxel as the centre of a peak. Assuming the peak must have a spherical profile, we count the total flux contained in spherical shells expanding away from the peak centre and consider the extent of the peak to end at the first shell where the total contained flux ceases to increase. We then search for the next-brightest peak excluding all voxels within this extent of the previous peak, repeating until we have identified the 200 brightest peaks in the [C\textsc{\,ii}] reconstruction. For comparison, we also identify the 200 most overdense peaks in LAE density contrast in an analogous fashion.

\begin{table}[]
    \centering
    \begin{ruledtabular}\begin{tabular}{rcc}
         & \multicolumn{2}{c}{Partner peaks}\\\cline{2-3}
         Data cube searched & 68\% interval & Extrema \\\hline
         WF reconstruction & (34, 51) & (33, 65) \\
         LAE cube & (52, 61) & (43, 64)\\
         LCB reconstruction & (61, 69) & (52, 76)
    \end{tabular}\end{ruledtabular}
    \caption{\added{68\% interval and minima/maxima across 10 realisations of the number of peaks identified (out of the brightest 200) in the [C\textsc{\,ii}] LIM signal that has a partner peak in either the reconstruction of the signal (through either the Wiener filter or the LCB filter) or the overlapping LAE density contrast cube (again identifying partner peaks from the brightest or most overdense 200 peaks in the mock observation used).}}
    \label{tab:partnerpeaks}
\end{table}

We then compare against the 200 brightest peaks identified analogously in the true [C\textsc{\,ii}] signal and see how many peaks have a partner peak identified in the reconstructions (or in the LAE density contrast cube), as defined by being within 10 voxels of the true peak. We tabulate the results in~\autoref{tab:partnerpeaks}. The Wiener-filtered data clearly skew towards identifying fewer partnered peaks while both the LAE data and the LCB reconstruction skew towards identifying more partnered peaks. By the metric, the LCB reconstruction somewhat outperforms the LAE cube, and significantly outperforms the WF reconstruction.

These results will vary significantly depending on the number of peaks we choose to identify in each cube, the definition of the peak extent and selection criterion, the smoothing scale for preprocessing the data cubes, and so on. Therefore, what we demonstrate here should hardly be seen as the limit of what these data and reconstruction methods can do in recovering cosmic structure. Nonetheless, the overall outcome demonstrates on a preliminary level that the LCB reconstruction may truly provide an improvement in identifying cosmic web features compared to either the LIM dataset alone or its cross-correlation target alone (the LAE catalogue in this case).}


\bibliography{ms}

\end{document}